\documentclass[useAMS,usenatbib]{mn2e}
\usepackage{graphicx}
\usepackage{amsmath,amssymb}
\usepackage[dvipdfmx]{color}
\newcommand{\del}{\partial}

\newcommand{\THE}{{\boldsymbol{\theta}}}

\newcommand{\K}{{\boldsymbol{k}}}

\newcommand{\BF}{\begin{figure}\begin{center}}
\newcommand{\EF}{\end{center}\end{figure}}
\newcommand{\BE}{\begin{equation}}
\newcommand{\EE}{\end{equation}}
\newcommand{\BEA}{\begin{eqnarray}}
\newcommand{\EEA}{\end{eqnarray}}

\newcommand{\tr}{\textrm}
\newcommand{\IG}{\includegraphics}
\def\v#1{\boldsymbol #1}
\newcommand{\ms}{M_{\odot}}
\begin{document}
\title{Constraints on warm dark matter from weak lensing in anomalous quadruple lenses}
\author[Kaiki Taro Inoue, Ryuichi Takahashi, Tomo Takahashi, Tomoaki Ishiyama]
{Kaiki Taro Inoue$^1$\thanks{E-mail:kinoue@phys.kindai.ac.jp}, Ryuichi
Takahashi$^2$, Tomo Takahashi$^3$ and Tomoaki Ishiyama$^4$
\\
$^{1}$Department of Science and Engineering, 
Kinki University, Higashi-Osaka, Osaka, 577-8502, Japan 
\\
$^{2}$Faculty of Science and Technology, Hirosaki University, 3
Bunkyo-cho, 
Hirosaki, Aomori 036-8561, Japan
\\
$^{3}$Department of Physics, Saga University, 
Saga 840-8502, Japan
\\
$^{4}$Center for Computational Science, University of Tsukuba, 1-1-1, Tennodai, Tsukuba, Ibaraki 305-8577, Japan
 }

%\author{Kaiki Taro Inoue\altaffilmark{1}, Ryuichi Takahashi \altaffilmark{2}}
%\altaffiltext{1}{Department of Science and Engineering, 
%Kinki University, Higashi-Osaka, Osaka 577-8502, Japan}
%\altaffiltext{2}{Faculty of Science and Technology, Hirosaki University, Hirosa%ki, Aomori 036-8561, Japan}

%%
%%
%%
%%
%%
\date{\today}

\pagerange{\pageref{firstpage}--\pageref{lastpage}} \pubyear{0000}

\maketitle

\label{firstpage}
\begin{abstract}
We investigate the weak lensing effect by line-of-sight structures with a surface 
mass density of $ \lesssim 10^8 \ms/\textrm{arcsec}^2$ in QSO-galaxy 
quadruple lens systems. Using high-resolution $N$-body simulations in
warm dark matter (WDM) models and observed four quadruple lenses 
that show anomalies in the flux ratios,  
we obtain constraints on the mass of thermal WDM, $m_{\textrm{WDM}}\ge
 1.3\,\textrm{keV}(95\%\textrm{CL})$ assuming that the density
of the primary lens is described by a singular isothermal
ellipsoid (SIE). The obtained constraint is consistent 
with those from Lyman-$\alpha$ forests and the
number counts of high-redshift galaxies at $z>4$. 
Our results show that WDM with a free-streaming comoving wavenumber
$k_{\tr{fs}} \le  27\, h/\textrm{Mpc}$ is disfavoured as the major component
 of cosmological density at redshifts $0.5 \lesssim z \lesssim 4$
provided that the SIE models describe the gravitational potentials of 
the primary lenses correctly.
\end{abstract}

\section{Introduction}
The clustering property of dark haloes at spatial scales
of $\lesssim 1\,\textrm{Mpc}$ is far from being understood.
In particular, the number of satellite galaxies in our Galaxy is by far
smaller than expected from theory, which is the so-called ''missing
satellite problem.'' As a solution, we may consider:
(1) baryonic solution - the star formation in the satellite galaxy
is suppressed due to some baryonic process. (2) dark matter solution -
a number of satellite galaxies are suppressed due to a large
free-streaming scale of dark matter particles. 

It has been known that the flux ratios of lensed images in some quadruply lensed
QSOs disagree with the prediction of best-fitting lens models with a smooth 
potential whose fluctuation scale is larger than the separation between the
lensed images. Such a discrepancy is called the 'anomalous flux ratio' and 
has been considered as an imprint of cold dark matter (CDM) subhaloes
with a mass of $\sim 10^{8-9} \ms$ in the lensing
galaxy \citep{mao1998,metcalf2001,chiba2002,dalal-kochanek2002,
metcalf2004,chiba2005,sugai2007,mckean2007,
more2009,minezaki2009}. 

However, recent studies based on high resolution simulations 
suggested that the predicted substructure population is too low 
to explain the observed anomalous flux ratios
\citep{amara2006, maccio2006,chen2009, xu2009,xu2010,chen2011}.
Moreover, the formation of dark satellites in lensing galaxies can be 
suppressed by baryonic processes, such as tidal stripping and outflows
due to supernovae. If the number density of satellites in our Galaxy
represents a typical value, the surface mass of dark satellites in lensing
galaxies should also be smaller than the expected values obtained from $N$-body 
simulations.  

Intergalactic haloes that are not belonging to lensing galaxies may evade
a suppression due to baryonic processes. Moreover, the lensing effects due to
line-of-sight haloes may play an important role \citep{chen2003,metcalf2005a,xu2012}. Intergalactic non-linear structures such as voids, walls, and filaments could also influence the flux ratios significantly. 

Indeed, taking into account astrometric shifts, recent studies 
have found that the observed anomalous flux ratios can be explained
solely by these line-of-sight structures with surface mass density $\sim 
10^{7-8}\ms/\textrm{arcsec}^2$ \citep{inoue-takahashi2012,
takahashi-inoue2014} without dark subhaloes 
in the lensing galaxies taken into account. The observed 
increase in the amplitude of  
magnification perturbations with redshift strongly implies that the origin 
is associated with sources rather than lenses.
If this is the case, one does not need to 
care about the suppression of dark satellites in the lensing galaxy due
to baryonic processes as a number of minihaloes in the line of sight are
not belonging to massive galaxies. 

Another mechanism that can suppress the number of dwarf galaxies is the
free-streaming of dark matter particles. If the thermal velocity at the
decoupling from the thermal bath is large enough, 
dark matter particles would erase the primordial
fluctuations at scales of dwarf galaxies.
Warm dark matter (WDM) particles
%sterile neutrinos, and superWIMPs 
are candidates for achieving such 
suppression.

However, if the suppression is too strong, the amount of neutral hydrogen
such as Lyman-$\alpha$ clouds is also significantly reduced. In fact,  
the best constraint on the mass scale of WDM comes from the 
observations of Lyman-$\alpha$ forests
\citep{viel2005, seljak2006, boyarsky2009, viel2013}.   
 
In a similar manner, one can constrain the mass or the
free-streaming scale of dark matter particles using anomalous 
quadruple lenses \citep{miranda2007}.
If the free-streaming scale is too large, or equivalently, 
the particle mass is too small, the amplitude of fluctuations of a surface mass
density in the line of sight becomes so small that the weak gravitational 
lensing effect, which acts as a perturbation to the flux ratios,  
becomes negligible. Therefore, the observed anomalous
flux ratios cannot be explained in such dark matter models.

In this paper, we revisit the weak lensing effect by the line-of-sight
structures in WDM models taking into account 
two important non-linear effects that have been overlooked in the
literature. One is the quick regeneration of the suppressed power of WDM models 
and the catching up with the linear and non-linear power of the CDM 
models \citep{boehm2005,schneider2012}. 
This effect might make WDM models difficult to 
exclude using QSO-galaxy lensing systems. Another is
the weak lensing effect due to non-linear objects such as walls, voids, 
and filaments. In CDM models, it turned out that the weak lensing effect
from locally underdense regions is also 
important for estimating magnification perturbation by the 
line-of-sight structures \citep{takahashi-inoue2014}. The weak lensing 
effect due to walls and filaments could also be important.
Therefore, we need to incorporate lensing effects due to non-linear objects 
in WDM models as well. For simplicity, however, we do not consider
lensing effects due to subhaloes in the lensing galaxies.

To take into account such non-linear effects, we first
calculate the non-linear power spectra of matter fluctuations down to 
mass scales of $\sim 10^5\, h^{-1 } \ms$ using $N$-body simulations. 
For simplicity, we do not consider baryonic effects 
in our simulations. Then we estimate the probability 
distribution of magnification perturbation for each lens using the
semi-analytic formulae developed in \citet{takahashi-inoue2014}.
We also take into account the astrometric shifts due to line-of-sight 
structures, which are often overlooked in the literature.

In Section 2, we describe our semi-analytic formulation for 
calculating the magnification perturbation due to line-of-sight structures. 
In Section 3, we show the results of our $N$-body simulations 
and the obtained non-linear power spectra in WDM models.  
In Section 4, we describe our samples of QSO-galaxy lensing
systems that show anomalies in the flux ratios. 
In Section 5, we present our results on the constraints on the mass of 
WDM particles and the free-streaming scales of dark matter particles.  
In Section 6, we conclude and discuss some relevant issues. 

In what follows, we assume a cosmology 
with a current matter density $\Omega_{m,0}=0.3134$, a baryon density 
$\Omega_{b,0}=0.0487$, a cosmological constant $\Omega_{\Lambda,0}=0.6866$,
a Hubble constant $H_0=67.3\, \textrm{km}/\textrm{s}/\textrm{Mpc}$,
a spectral index $n_s=0.9603$, and the root-mean-square (rms) 
amplitude of matter fluctuations at $8 h^{-1}\, \textrm{Mpc}$, 
$\sigma_8=0.8421$, which are obtained from the observed 
cosmic microwave background (Planck + WMAP polarization, \citet{ade2014}).

\section{semi-analytic formulation}
\begin{figure}
\hspace{-0.16cm}
\IG[width=85mm]{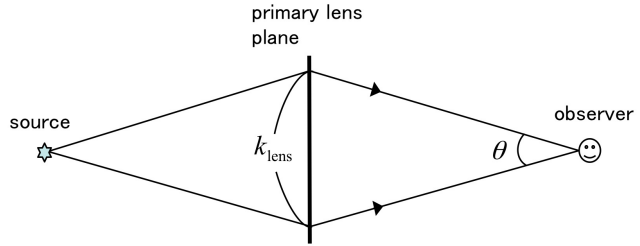}
\caption{Schematic diagram of unperturbed light rays. The wavenumber
 $k_{\tr{lens}}$ is defined as $k_{\tr{lens}}=\pi/2b$, where $b$ is the mean
 separation between lensed images and the centroid of the primary lens galaxy.  }
\label{weaklens}
\end{figure}
In this section, we briefly describe our 
semi-analytical formulation (for details, see 
\citet{inoue-takahashi2012,takahashi-inoue2014}). 

We use a statistic $\eta$ to measure the magnification perturbation of lensed images in 
QSO-galaxy lens systems:
\BE
  \eta \equiv \biggl[\frac{1}{2 N_{\rm pair}} \sum_{{\rm i} \neq {\rm j}} \left[
 \delta^\mu_{\rm i} ({\rm minimum}) - \delta^\mu_{\rm j} ({\rm saddle})
 \right]^2 \biggr]^{1/2},
\label{eta_def}
\EE
where $\delta^\mu({\rm minimum})$ and $\delta^\mu({\rm saddle})$ are
magnification (denoted by $\mu$ ) contrasts $\delta^\mu \equiv \delta \mu/\mu$ 
corresponding to the minimum and saddle images and $N_{\rm pair}$
denotes the number of pairs of lensed images. If the correlation
of magnification between pairs of images is 
negligible, then $\eta $ corresponds to the mean 
magnification perturbation of one of lensed images. For instance, 
$\eta=0.1$ means that the magnification is expected to change 
by $10$ percent. Note that we need to fix the primary lens model (i.e., 
a best-fittingted model without line-of-sight structures) 
in order to calculate $\eta$. In other words, $\eta$ is a model 
dependent statistic. 

The second moment of the magnification 
perturbation $\eta$ can be calculated as follows.
First, we need to estimate a perturbation $\varepsilon $ to
the largest angular separation $\theta_{\rm max} $ between
a pair of lensed images X and Y due to the line-of-sight structures,
\BE
\varepsilon =|\delta \v{\theta}(\rm{X})-\delta \v{\theta} (\rm{Y})|,
\EE
where $\delta \v{\theta}$ represents the astrometric shift perturbation 
of a lensed image at $\v{\theta}$ in the lens plane. 
We then assume that the perturbation satisfies $\varepsilon \le
\varepsilon_0$, where $\varepsilon_0$ is the observational error for the largest angular
separation. This condition gives an approximated upper limit on the 
contribution of line-of-sight structures assuming that the gravitational 
potential of the primary lens is sufficiently smooth on the scale of the
Einstein radius and the projected density has a nearly-circular elliptical 
symmetry (see appendix in \citet{takahashi-inoue2014}).  

In order to satisfy such a condition, we assume
that small-scale modes with a wavelength larger than  
the mean comoving separation $b$ between the lens centre
and lensed images at the primary lens plane are significantly suppressed. 
Any modes whose fluctuation scales are larger than
$b$, which is roughly the size of the comoving Einstein radius, 
contribute to the smooth component of a 
primary lens, namely, a constant convergence and shear (Fig. \ref{weaklens}). 
Therefore, we consider only modes whose wavenumbers satisfy $k>k_{\rm
lens}$ where $k_{\rm lens} \equiv \pi/2b$. 
Otherwise, double-counting of the constant convergence and shear
leads to a systematically large perturbation.
Furthermore, we also assume that modes with wavenumbers
$k_{\rm lens}<k<k_{\rm cut}$ are suppressed to some extent. 
These modes correspond to the secondary lens [modelled by a singular
isothermal sphere (SIS) or singular isothermal ellipsoid (SIE)] 
in the line of sight. In our samples, MG0414+0534, B1608+656 and
B2045+265 have the secondary lens. In our sample, 
the lensing galaxies of these systems ($k_{\tr{lens}}=100\sim 200 h/\tr{Mpc}$) are typically more massive than those without
a secondary lens ($k_{\tr{lens}}=400\sim 500 h/\tr{Mpc}$). The cut off scale $k_{\rm cut}$ is determined
by the condition that the perturbation $\varepsilon$ 
of an angular separation $\theta$ between an arbitrary 
pair of lensed images should not exceed the observational error
$\varepsilon_{\rm{obs}}$ for the maximum separation angle between lensed images.

In what follows, we adopt a filtering the so-called constant shift
(CS) filter \citep{takahashi-inoue2014},
\BE
  W_{\rm{CS}}(k;k_{\rm cut}) = \left\{ 
\begin{array}{ll}
W_{\rm{int}}(k), & \mbox{$k<k_{\rm cut}$} \\
1, & \mbox{$k \ge k_{\rm cut}$},
\end{array}
\right.
\EE
in which the corresponding contribution to the angular shifts between
a pair of images with the maximum separation angle $\theta_{\max}$ is constant
in logarithmic interval in $k$ for $k<k_{\rm cut}$. The CS filter
mildly suppresses the large angular-scale modes 
with wavenumbers $k_{\rm lens}<k<k_{\rm cut}$ by keeping 
the contribution to an angular shift $\varepsilon$ constant in $\log{k}$
and gives a relatively good approximation in the CDM models \citep{takahashi-inoue2014}.

$W_{\rm{int}}$ is explicitly given by
\BE
W^2_{\rm{int}}(k;k_{\rm cut})\equiv \frac{\frac{\displaystyle \del
\langle \varepsilon^2 \rangle }{\displaystyle \del \ln{k}}\bigg |_{k=k_{\rm cut}}}
{\frac{\displaystyle\del\langle \varepsilon^2 \rangle }{\displaystyle\del \ln{k}}},
\EE
where
\BE
\langle \varepsilon^2 \rangle =2 \langle \delta\theta^2 (0) \rangle - 
2 \langle \delta\theta (0)\delta \theta(\theta_{\rm{max}}) \rangle, 
\EE
and
\BEA
\langle \delta \theta (0) \delta \theta (\theta) \rangle 
&=&
\frac{9 H_0^4 \Omega_{m,0}^2}{ c^4}
\int_0^{r_S} dr  \biggl(\frac{r-r_S}{r_S} \biggr)^2 [1+z(r)]^2
\nonumber
\\
&\times& \int_{k_{\rm {lens}}}^{\infty}\frac{dk}{2 \pi k} W_{\textrm{CS}}^2(k;k_{\textrm{cut}})
P_{\delta}(k, r) J_0(g(r) k\theta),
\nonumber
\\
\label{eq:shiftsq}
\EEA
where
\BE
g(r)= \left\{ 
\begin{array}{ll}
r, & \mbox{$r<r_L$} \\
{r_L(r_S-r)}/{(r_S-r_L)}, & \mbox{$r\ge r_L$}
\end{array}
\right.
\label{eq:g}
\EE
and $P_\delta(k,r)$ is the power spectrum of dark matter 
density fluctuations as a function of the wavenumber $k$ and the
comoving distance $r$. $r_S$ is the comoving distance to the source
and $r_L$ to the lens from an observer and $z(r)$ is the redshift of a 
point at a comoving distance $r$. $\langle \rangle$ represents
an ensemble average. $J_0$ is the zero-th order Bessel function.
$g(r) \theta$ denotes the tangential separation between two unperturbed
light rays at a comoving distance $r$ from the observer.

Once $k_{\rm{lens}}$ and $k_{\rm{cut}}$ are determined, we can
compute the constrained perturbed convergence $\delta \kappa$ and shear $\delta
 \gamma_{1,2}$ as functions of a separation angle $\THE$ between
a pair of lensed images. For instance, the 
constrained two-point correlation of $\delta \kappa$ as a
function of a separation angle $\THE$ is 
\BEA
\xi_{\kappa \kappa}(\THE) &\equiv&
\langle \delta \kappa (0) \delta \kappa (\THE) \rangle  \nonumber \\
&=&\frac{9 H_0^4 \Omega_{m,0}^2}{4 c^4}
\int_0^{r_S} dr  r^2 \biggl(\frac{r-r_S}{r_S} \biggr)^2 [1+z(r)]^2
\nonumber \\
&&\times \int_{k_{\rm {lens}}}^{\infty}\frac{dk}{2 \pi} k
 W_{\textrm{CS}}^2(k;k_{\textrm{cut}})  P_{\delta}(k,r)
 J_0(g(r) k\theta).  \nonumber \\
\label{eq:ck}
\EEA
To calculate $P_\delta$, we use a
fitting function obtained from high resolution
cosmological simulations (see also
 \citet{smith2003, inoue-takahashi2012, takahashi2012, takahashi-inoue2014}).
The fitting function for the WDM model can be used up to a wavenumber 
 $k \sim 300\,h {\rm Mpc}^{-1}$ at $0 \leq z \leq 3$ within $\sim 20\%$ accuracy 
(see Section 3).

The two-point correlation functions for the other perturbed quantities
are obtained by the following substitution in equation (\ref{eq:ck}): 
\BEA
 \langle \delta \gamma_1(0) \delta \gamma_1(\THE) \rangle &:& J_0 \rightarrow
  \frac{1}{2} \left[ J_0 + J_4 \cos(4 \phi_\theta) \right], \nonumber \\
 \langle \delta \gamma_2(0) \delta \gamma_2(\THE) \rangle &:& J_0 \rightarrow
  \frac{1}{2} \left[ J_0 - J_4 \cos(4 \phi_\theta) \right], \nonumber \\
 \langle \delta \kappa(0) \delta \gamma_1(\THE) \rangle &:& J_0 \rightarrow
  - J_2 \cos(2 \phi_\theta), \nonumber \\
 \langle \delta \kappa(0) \delta \gamma_2(\THE) \rangle &:& J_0 \rightarrow
  - J_2 \sin(2 \phi_\theta), \nonumber \\
 \langle \delta \gamma_1(0) \delta \gamma_2(\THE) \rangle &:& J_0 \rightarrow
  \frac{1}{2} J_4 \sin(4 \phi_\theta), \label{other_corr}
\EEA
where $\THE=(\theta \cos \phi_\theta, \theta \sin \phi_\theta)$ and
the Bessel functions $J_{0,2,4}$ are functions of $g(r)k\theta$.
From equations (\ref{eta_def}), (\ref{eq:ck}) and (\ref{other_corr}) , we can 
obtain the second moment of $\eta$.  

For example, let us consider three images with two minima A and C and one saddle B with $\kappa_B<1$. 
Choosing coordinates where the separation angle is perpendicular to $+$
mode (i.e., $\theta \sin \phi_\theta=0$), we have $\langle \delta \kappa \delta \gamma_2
 \rangle = \langle \delta \gamma_1 \delta \gamma_2
 \rangle =0$. Then, for $|\delta_i^\mu|\ll1$, the second moment $\langle \eta^2 \rangle$ can
 be written as
\BEA
\langle \eta^2
\rangle&=&\frac{1}{4}\biggl[(I_A+I_B)-2I_{AB}(\theta_{AB})+(I_B+I_C)
\nonumber
\\
&-& \!\!\!\! 2I_{BC}(\theta_{BC}) \biggr],
\label{eq:estimator-anal}
\EEA
where
\BE
I_i\equiv \mu_i^2(4(1-\kappa_i)^2+2 \gamma_{1i}^2 + 2\gamma_{2i}^2)
\xi_{\kappa}(0) ,
\label{eq:Ii}
\EE
and
\BEA
I_{ij}(\theta)\!\!&\equiv&\!\!
4 \mu_i \mu_j \biggl[(1-\kappa_i)(1-\kappa_j)\xi_\kappa(\theta)
\nonumber
\\
&+&
\gamma_{1i}\gamma_{1j}\langle \delta \gamma_1 (0) \delta \gamma_1
(\theta) \rangle
+\gamma_{2i}\gamma_{2j}\langle 
\delta \gamma_2 (0) \delta \gamma_2(\theta) \rangle
\nonumber
\\
&+&
(1-\kappa_i)\gamma_{1j} \langle \delta \kappa_i(0) \delta \gamma_{1j}
(\theta) \rangle
\nonumber
\\
&+&
(1-\kappa_j)\gamma_{1i} \langle \delta \kappa_j(0) \delta \gamma_{1i}
(\theta) \rangle
\biggr],
\nonumber
\\
\label{eq:Iij}
\EEA
for $i=\rm{A,B,C}$.  
In a similar manner, for a four-image system with 
two minima A and C and two saddles B and D with $\kappa_{\textrm{B}}<1$
and $\kappa_{\textrm{D}}<1$, the second moment is given by
\BEA
\langle \eta^2 \rangle&=&\frac{1}{8}
\biggl[I_A+I_B 
-2I_{AB}(\theta_{AB}) + (I_C+I_B)
\nonumber
\\
&-&\!\!2I_{CB}(\theta_{CB})+(I_A+I_D) 
-2I_{AD}(\theta_{AD})
\nonumber
\\
&+&\!\!(I_C+I_D)
-2I_{CD}(\theta_{CD})
\biggr],
\nonumber
\\
\EEA
where $I_i$ and $I_{ij}(\theta)$, $i=\rm{A,B,C,D}$ are given by
equations (\ref{eq:Ii}) and (\ref{eq:Iij}).
Note that we are using coordinates in which $\phi_\theta=0$.

\section{Non-linear power spectrum}

\subsection{Initial condition}
We calculate the initial power spectrum in models with WDM by using the modified version of CAMB 
\citep{camb}. We assume thermal distribution for WDM and all dark matter component 
being WDM. Since we fix the abundance of WDM, its mass $m_{\rm WDM}$ and the 
temperature of WDM species $T_{\rm WDM}$ are related as 
\begin{equation}
\label{eq:omega_WDM}
\Omega_{\rm WDM} h^2 = \left( \frac{T_{\rm WDM}}{T_\nu} \right)^3 \left( \frac{m_{\rm WDM}}{94~{\rm eV}} \right),
\end{equation}
where $T_\nu$ is the temperature of neutrinos.  By the effect of the free-streaming of WDM particles,
the cosmic structure can be erased and the matter power spectrum damps on small scales, which is commonly characterized 
by the free-streaming scale $\lambda_{\rm fs}$, defined by the comoving length that 
WDM particles free-stream until the radiation-matter equality time. 
$\lambda_{\rm fs}$ is explicitly given by \citep{kolb_turner}
\BEA
\label{eq:lambda_fs}
\lambda_{\rm fs}& = & 0.114~ {\rm Mpc}
\left( \frac{1~{\rm keV}}{m_{\rm WDM}} \right)
\left( \frac{10.75}{g_* ( T_D) } \right)^{1/3}
\nonumber
\\
&\times& \left[ 2 + \log \left( \frac{t_{\rm eq}}{t_{\rm NR}} \right) \right],
\EEA
where $t_{\rm eq}$ and $t_{\rm NR}$ are the time of radiation-matter 
equality and that when WDM particles become non-relativistic, respectively.
$g_\ast (T_D)$ is the effective number of degrees of freedom at the time of decoupling of WDM particles, 
denoted by the temperature $T=T_D$.
In the following analysis, we fix the energy density of WDM as 
$ \Omega_{\rm WDM} = 0.2647$; hence, the temperature $T_{\rm WDM}$ (or $g_\ast (T_D)$) and 
the mass $m_{\rm WDM}$ are related through equation ~\eqref{eq:omega_WDM}. 

Above arguments are valid for thermally produced WDM species. However, other candidates for WDM 
such as sterile neutrinos \citep{Dodelson:1993je} have also been discussed in the literature. 
For the sterile neutrinos produced via active-sterile neutrino oscillations, 
its distribution function can be approximated by a generalized Fermi-Dirac distribution,
then the effect of sterile neutrino can be regarded as the same as the one for WDM
by using the following identification for the mass 
\citep{Colombi1996, viel2005}:
\begin{equation}
%\label{ }
m_\textrm{s} = 4.46~{\rm keV} \left( \frac{m_{\rm WDM}}{1~{\rm keV}} \right)^{4/3} \left( \frac{0.12}{\Omega_{\rm WDM}h^2} \right)^{1/3}.
\end{equation}
From this formula, one can derive the constraint for the mass of sterile
neutrino once we obtain that for thermally produced WDM.

\begin{figure}
\hspace{-0.26cm}
\IG[width=85mm]{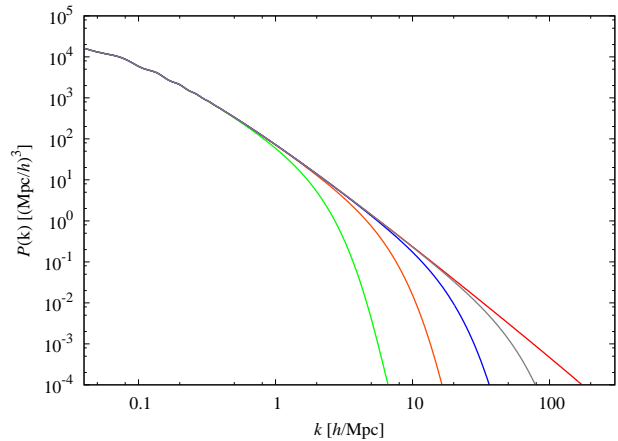}
\caption{Plots of linear matter power spectra for the $\Lambda$CDM model (red solid line), 
WDM models with $k_{\rm fs}=2 \upi/\lambda_{\rm fs} =140\,h$/Mpc (grey), $44\,h$/Mpc (blue),
 $15\,h$/Mpc (orange) and $4.8\,h$/Mpc (green).}
\label{fig:Pk}
\end{figure}

In Fig.~\ref{fig:Pk}, we show the linear matter power spectra in the
$\Lambda$CDM model, and WDM models 
with $k_{\rm fs}=2 \upi/\lambda_{\rm fs} = 140, 44, 15$ and $4.8\,h$/Mpc. 
The corresponding WDM masses are listed in Table~\ref{table_cdm_wdm}.

\subsection{$N$-body simulation}

We run cosmological $N$-body simulations to investigate the non-linear matter
 power spectra of WDM models.
Our purpose is to obtain the fitting formula of non-linear power
 spectra used in our analytical formula (see Section 2).
In order to cover a wide-range scale of gravitational evolution, we 
 run simulations with two different boxes with a side of $100h^{-1}$Mpc and $10h^{-1}$Mpc,
 hereinafter referred to as $L100$ and $L10$, respectively.
The number of particles in the boxes is set to $1024^3$.
%Here, since the small-box size is very small ($L=10h^{-1}{\rm Mpc}$ on a
% side), the fluctuations larger than the box-size are not included 
% comparable to  
%Thus, we use the mode $1/10$ times smaller than the box-size,
% which is corresponding to $k>60h/$Mpc.
The initial positions and velocities of particles are given at redshift $z_{\rm init}=24$
based on second-order Lagrangian perturbation theory \citep{crocce2006,nishimichi2009}.
We adopt a concordant CDM model and four WDM models with free-streaming wavenumbers 
 $k_{\rm fs}=2 \upi/\lambda_{\rm fs} = 140, 44, 15$ and $4.8\,h$/Mpc in our simulations.
The CDM and WDM models are summarized in Table \ref{table_cdm_wdm}. 
The input linear power spectra of the CDM and WDM models are evaluated using
 CAMB (see Section 3.1). In our simulations, we ignore the thermal
 motion of WDM particles, which can be verified as follows.
 The rms thermal velocity of WDM particles at the initial redshift
 ($z_{\rm init}=24$) is $\sigma_v \simeq 1.1 {\rm km/s} (g_{\rm WDM}/1.5)^{1/3}
 (m_{\rm WDM}/{\rm k\,eV})^{-4/3}$ in our cosmological model,
 where $g_{\rm WDM}$ is the degree of freedom of the WDM particle
 \citep{bode2001}. 
On the other hand, the rms physical peculiar velocity of the particles at the
initial time is $\gtrsim 10\,$km/s in our WDM models.
Thus, we can ignore the thermal motion of WDM particles (see also similar
 discussion in \cite{angulo2013}). 

To follow the gravitational evolution of the dark matter particles,
 we employ publicly available tree-PM codes, Gadget2
 \citep{springel2001,springel2005} for
 the large-box simulation (L100) and GreeM \citep{ishiyama2009,ishiyama2012}
 for the small-box simulation (L10).
GreeM is tuned to accelerate the tree gravitational calculation, and it
 is faster than Gadget2 especially in the strongly non-linear regime.
Hence, we employ GreeM for the small-box simulation. 
The PM meshes are $2048^3 (512^3)$ for the L100 (L10).
The particle Nyquist wavenumbers are $k_{\rm Nyq}=32.2 (322) h/$Mpc for
 the L100 (L10).   
The gravitational softening length is set to $3\%$ of the mean particle
 separation.
The simulation snapshots are dumped at redshifts $z=0, 0.3, 0.6, 1, 2$ and $3$.
We prepare $3 (5)$ independent realizations for the L100 (L10)
 for each CDM or WDM model to reduce the sample variance.
Our simulation settings are summarized in Table \ref{table_sim}.

We check the accuracy of our simulation results as follows.
For Gadget2, we use the same simulation parameters (time step, force
 accuracy and so on) in \citet{takahashi2012} (Section 2) in which we
 achieved a few percent accuracy of the power spectra.
% checked the code using $L=320h^{-1}$Mpc with $1024^3$ particles
For GreeM, we run simulations with finer simulation parameters and confirmed
 that the power spectra have $<1\%$ accuracy for $k<300h$/Mpc. 

To evaluate the matter power spectra from the particle distribution,
 we assign $1024^3$ particles into
 $1536^3$ grids using the Cloud-in-Cell (CIC) method \citep{hockney1988} to
 obtain the density fluctuations.
Then, we perform FFT\footnote{FFTW home page: http://www.fftw.org/} and
 calculate the power spectrum:
\BE
  P(k) = \frac{1}{N_k} \sum_{\K^\prime}
 \left| \tilde{\delta}(\K^\prime) \right|^2,
\EE
where the summation is done over a range of $k-\Delta k/2 < |\K^\prime|
 < k+\Delta k/2$ with a bin-width $\Delta k$, and $N_k$ is the number of modes
 in a $k$ bin. 
We also employ the holding method \citep[e.g.][]{jenkins1998,smith2003} 
 to probe smaller scales.
We calculate the mean power spectra and $1\sigma$ errors from 
 5(3) realizations in the L100(L10).

\begin{figure}
%\epsscale{0.9}
\vspace*{1.0cm}
\includegraphics[width=85mm]{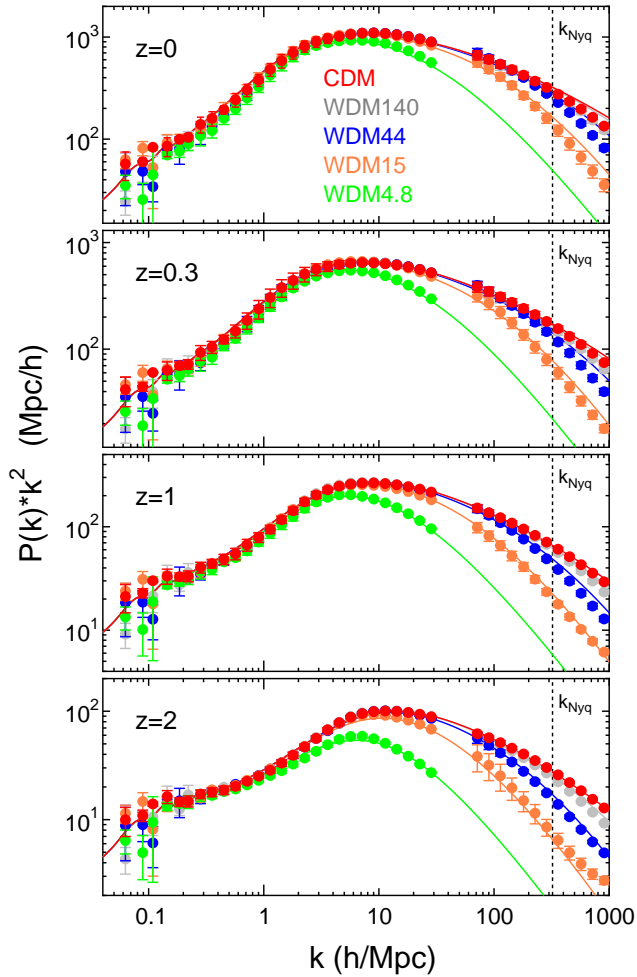}
\caption{
Non-linear matter power spectra for CDM and WDM models
 for various redshifts $z=0, 0.3, 1$ and $2$.
The filled circles with the error bars are the simulation results
 for CDM(red), WDM140(grey), WDM44(blue), WDM15(orange) and WDM4.8(green) in
 Table \ref{table_cdm_wdm}.
The results are taken from the large-box simulations (L100) for $k<30h$/Mpc and 
 the small-box simulations (L10) for $k>60h$/Mpc.
Note that the vertical axis is $P(k) k^2$ (not $P(k)$) to show the
 differences among the models clearly.
The vertical dotted line denotes the Nyquist wavenumber of the small-box
 simulations (L10).
The solid curves are obtained from our fitting formula for the WDM
 models (see the main text and Appendix A). 
}
\label{fig_pk}
\vspace*{0.5cm}
\end{figure}

Fig.\ref{fig_pk} shows our simulation results for the matter power spectra
in the CDM and WDM models shown in Table \ref{table_cdm_wdm} at redshifts
 $z=0,0.3,1$ and $2$. 
The filled circles with error bars are the mean power spectra with 
the errors obtained from the realizations of simulations.
The results are taken from the large-box simulations (L100) for
 $k<30h$/Mpc and from the small-box simulations (L10) for $k>60h$/Mpc.
Here, $k=30h$/Mpc is the Nyquist wavenumber of the L100, and $k=60h$/Mpc
 corresponds to a scale of $1/10$ times smaller than the small
 box-size (L10)\footnote{The box-size of the L10 is very small
 ($L=10h^{-1}$Mpc on a side) and hence it does not include density
 fluctuations larger than the box size that may affect small-scale
 clustering via mode coupling. To avoid this, we use only modes much
 smaller than the box size.}.
The vertical dotted line denotes the Nyquist wavenumber of the small-box
 simulation (L10).
The solid curves are obtained from our fitting formula based on the halofit 
 model for a $\Lambda$CDM model \citep{smith2003,takahashi2012}, but
 slightly modified in WDM models. Our formula is based on  
the numerically obtained power spectra with the maximum 
wavenumber $k_{\rm max}=300 (30) h /{\rm Mpc}$
 for the L10(L100). Details of the model fitting 
parameters are given in Appendix A. The simulation box-size should be 
much larger than the free-streaming scales in WDM models to follow 
gravitational evolution accurately.
Thus, we do not use the simulation results for the WDM4.8 in the
 small-box simulation (L10), because
 its free-streaming scale ($\lambda_{\rm fs}= 2 \upi/k_{\rm fs} =1.3h^{-1}$Mpc)
 is close to the box size ($10h^{-1}$Mpc).
As shown in Fig. \ref{fig_pk}, the suppression due to free-streaming of WDM particles
 becomes less prominent at low 
redshifts even though the initial power spectra of the WDM models are
 exponentially suppressed at small scales $k \gtrsim k_{\rm fs}$.
For example, the initial power spectrum $P(k)$ of the WDM15 is ten orders of magnitude
 smaller than that of the CDM at $k=300\,h/$Mpc, but the ratios become
 only $\sim 2 (4)$ at low redshifts $z=0 (2)$.
This result exhibits power transfer from large to small scales
 via the mode coupling during the non-linear evolution 
\citep{bp1997,wc2000,sm2011,viel2012}. 
The quick regeneration of the suppressed power of WDM models 
and catching up with the linear and non-linear power of the CDM
play an important role for estimating the lensing effects due to 
line-of-sight structures. The small-scale powers 
with wavenumbers $k> 300\, h/$Mpc 
may be systematically larger than the extrapolated values due to 
spurious fragmentation of filaments at the scale of grids \citep{wang2007}.
However, for fitting, we use only 
power spectra at scales larger than $k=300\,
h/\tr{Mpc}$ where the numerical convergence is confirmed. 
Therefore, such numerical noises do not affect our estimates (see Appendix A). 
From our simulations with $512^3$ and $1024^3$ particles, it turns out
that for $k> 320
h/\tr{Mpc}$,  the error of $P(k)$ can be estimated as $\sim 10\%$ and
our fitting formula overestimates the simulation results
by a factor of $1\sim 2$ for $320<k< 1000 h/\tr{Mpc}$. 
In what follows, we use $P(k)$ obtained from our fitting formula,
which would yield a conservative constraint on the WDM mass.

\begin{table}
\hspace{-1.5cm}
%\begin{minipage}{170mm}
\caption{CDM and WDM models}
\setlength{\tabcolsep}{2pt}
\begin{tabular}{lccccc}
\hline
\hline
  Model & $k_{\rm fs}(h {\rm Mpc}^{-1})$ & $m_{\rm WDM}({\rm keV})$ \\ 
\hline
 CDM & $---$ & $---$  \\
\hline
 WDM140 & $140$ & 5.0 \\
\hline
 WDM44 & $44$ & 1.9 \\
\hline
 WDM15 & $15$ & 0.77 \\
\hline
 WDM4.8  & $4.8$ &  0.29 \\
\hline 
\label{table_cdm_wdm}
\end{tabular}
\\
 Note: The CDM and WDM models in our simulations. We show the
 free-streaming wavenumbers $k_{\rm fs}$
 and WDM particle masses $m_{\rm WDM}$.  
%\end{minipage}
\end{table}

\begin{table*}
\hspace{-1.5cm}
%\begin{minipage}{170mm}
\caption{Our simulation setting}
\setlength{\tabcolsep}{2pt}
\begin{tabular}{lcccccc}
\hline
\hline
   & $L(h^{-1} \rm{Mpc})$ & $N_{\rm p}^3$ &
 $k_{\rm Nyq} (h \rm{Mpc}^{-1})$ & $m_{\rm p} (h^{-1} M_\odot)$ &
 $z$ & $N_{\rm r}$  \\ 
\hline
 L100 & $100$ & $1024^3$ & $32.2$ & $8.1 \times 10^7$ & $0,0.3,0.6,1,2,3$
 & $3$ \\
\hline
 L10 & $10$ & $1024^3$ & $322$ & $8.1 \times 10^4$ & $0,0.3,0.6,1,2,3$
 & $5$ \\
\hline 
\label{table_sim}
\end{tabular}
\\
\flushleft{
Note: Parameters in our simulations are side length of simulation box $L$, number of dark
 matter particles $N_{\rm p}^3$, Nyquist wavenumber $k_{\rm Nyq}$,
 particle mass $m_{\rm p}$, redshifts of the simulation outputs $z$
 and number of realizations $N_{\rm r}$.}
%\end{minipage}
\end{table*}

\section{Lens model}
As a fiducial model of lensing galaxies, we adopt an SIE \citep{kormann1994}, which can explain
flat rotation curves. We use the fluxes of lensed images, the relative positions of lensed quadruple
images and the centroid of lensing galaxies and time delay of lensed
images if available. The contribution from groups, clusters, and large-scale
structures at angular scales larger than the
Einstein radius of the primary lens is taken into account as an external
shear (ES). The parameters of the SIE('s) plus ES model 
are the angular scale of the critical curve or the 
mass scale inside the critical
curve $b'$, the apparent ellipticity $e$ of the lens and its
position angle $\theta_e$, the strength and the direction of the ES
$(\gamma,\theta_\gamma)$, the lens position $(x_{G},y_{G})$ and the source
position $(x_s,y_s)$. The Hubble constant $h$ is also treated as a model
parameter. The angles $\theta_e$ and $\theta_\gamma$ are measured in
East of North expressed in the observer's coordinates (see Table 4). 

To find a set of best-fitting parameters,
we use a numerical code called GRAVLENS \footnote{See
http://redfive.rutgers.edu/$\sim$keeton/gravlens/} developed by Keeton in order to implement
the simultaneous $\chi^2$ fitting of the fluxes, positions, and time delay of lensed
images (if reliable data are available) 
and the positions of centroid of lensing galaxies.
The total $\chi^2_{\tr{tot}}$ is equal to the sum of $\chi^2_{\tr{imag}}$ for lensed
images, $\chi^2_{\tr{flux}}$ for fluxes,  $\chi^2_{\tr{tdel}}$ for time delays, 
and $\chi^2_{\tr{gal}}$ for the positions of lensing galaxies.
\section{Quadruple lens systems}
In the following, we shortly describe six quadruple lens systems
that show a large cusp relation $R_{\textrm{cusp}}$ or fold relation
$R_{\textrm{fold}}$\footnote{For instance, if image A and B are minima
and image C is saddle,
$R_{\tr{cusp}}=(\mu_A+\mu_C+\mu_B)/(\mu_A+|\mu_C|+\mu_B)$ for the cusp lenses
and $R_{\tr{fold}}=(\mu_A+\mu_C)/(\mu_A+|\mu_C|)$ or $(\mu_B+\mu_C)/(\mu_B+|\mu_C|)$ for the fold lenses.}.
It is known that eight quadruple lens systems show
apparent anomalies in the radio flux ratios (e.g.,\citet{ 
mao1998,metcalf2001,chiba2002,dalal-kochanek2002,metcalf2004}). 
However, we exclude B1555+375 and B1933+503 in our analysis 
since the redshifts of the lens and source of B1555+375
are not measured and the spiral lens B1933+503 has a very complex 
structure \citep{suyu2012}.
In our analysis, we use the observed MIR fluxes for MG0414+0534 and the 
radio fluxes averaged over a
certain period for other five systems. It should be noted that the MIR fluxes
are microlens free especially for high-redshift sources. 
For the astrometry, we use optical or
NIR data in order to avoid bias due to complex structures of jets. We
also use time delay for modelling B1608+656. 
We find that B2045+265 and B1608+656 with large $R_\textrm{cusp}\!\sim\! 0.5$ are no longer
anomalous ($\chi^2_{\tr{flux}} \le 1$ for each lensed image) 
if a companion galaxy G2 at the redshift of the primary lens G1 is
taken into account. Note that our result for B2045+265 is consistent with the
previous work \citep{mckean2007}.
Therefore, we use four anomalous quadruple lenses B1422+231, B0128+437,
MG0414+0534, and B0712+472 for constraining the mass of WDM particles. 
All the observed data used in our analysis are listed in table 3. 
In what follows, we describe the property of each lens.
\subsection{B1422+231}
The cusp-caustic lens B1422+231 consists of three bright images
A, B, C, and a faint image D. 
Images A and C are minima, and B and D are saddles.   
The quasar redshift $z_S=3.62$ is the largest in our four samples 
and the primary lensing galaxy is at $z_L=0.34$
\citep{kundic1997b,tonry1998} and the measured ellipticity and the 
position angle of near-infrared (near-IR) light distribution are $e=0.39\pm 0.02$ and 
$\theta_e=-58.90\pm0.80(^\circ)$ \citep{sluse2012}.
We use the radio flux ratios \citep{koopmans2003} of four images at 5 GHz
averaged over a period of 8.5 months, which are
consistent with the MIR counterparts \citep{chiba2002}. 
We also use the astrometry of lensed images and the centroid of the primary
lensing galaxy in \citet{sluse2012} obtained from the use of Magain-Courbin-Sohy (MCS) 
deconvolution algorithm applied in an iterative way 
(ISMCS) to near-IR Hubble Space Telescope (HST) images. The maximum total error in
the positions of lensed images is 1.05\,mas. Therefore, we
assume an error of $\sqrt{2}\times 1.05\sim 1.4\,$mas 
for the angular separations of 
lensed images. The positions of
lensed images and the centroid G of the primary lensing galaxy are well
fitted by an SIE and an ES assuming that the error in the
angular position of G is 0.01 arcsec. However, the flux ratios are
not well fitted. We find that addition of lensing potential with 
low-multipole terms  ($m=3$ or $m=4$) or changing the power index of
radial profile in the mass density does not improve the fit.
\citet{chiba2002} and \citet{nierenberg2014} argue a presence of
substructure around A. Alternatively, the possible perturber may be a
halo or some other objects in the line of sight. For computing 
the magnification perturbation, we use only three bright images
as the signal-to-noise ratio of D/B is significantly worse than
the other images. 
\subsection{B0128+437}

The fold-caustic lens B0128+437 consists of one bright image A, and 
three fainter images B, C and D. The images A and C are minima, and B and D are
saddles. The quasar redshift is $z_S=3.124$ \citep{mckean2004} 
and that of the primary 
lensing galaxy is either $z_L=0.645$ or $1.145$ \citep{lagattuta2010}. 
Combining with the previous photometric and spectroscopic data, the
latter choice is favoured than the former \citep{mckean2004, 
lagattuta2010}. Therefore, we assume $z_l=1.145$ in what follows.  
We use the radio flux ratios \citep{koopmans2003} of four images at 5 GHz
averaged over a period of 8.5 months, and the astrometry in 
\citet{lagattuta2010} obtained from 
ground-based near-IR imaging coupled with laser guide-star adaptive
optics. We also assume that the astrometric errors of each lensed 
image are $0.005$ arcsec \citep{lagattuta2010}.
Although the positions of lensed images and the centroid of the primary
lensing galaxy G can be fitted by an SIE plus an ES, the
predicted flux ratios show discrepancy with the data. There might be 
a sub/line-of-sight halo around C. 
\subsection{MG0414+0534}
The fold-caustic lens MG0414+0534 consists of two bright images A1 and A2, and 
two faint images B and C. The images A1 and B are minima, and A2 and C are saddles.
A source quasar at $z_S=2.64$ is lensed by an elliptical galaxy at
 $z_L=0.96$ \citep{hewitt1992,lawrence1995, tonry1999}.
A simple lens model, an SIE with
an external shear (SIE-ES) cannot fit the image positions as well as 
the flux ratios. \citet{schechter1993} and \citet{Ros2000} 
suggested that another galaxy called ``X'' is necessary for
fitting the relative image positions.
We use the MIR flux ratios A2/A1 and B/A1 measured 
by \citet{minezaki2009} and \citet{macleod2013} since the 
radio fluxes might be hampered by Galactic refractive
scintillation\citep{koopmans2003}. For the astrometry, 
we use the data from the CASTLES (CfA-Arizona Space Telescope LEns Survey) 
data base of gravitational lens. Although, the positions are well
fitted by an SIE and an ES plus an SIS that accounts for object X, 
the flux ratios are not well fitted. A possible sub/line-of-sight halo 
near A2 significantly improves the fit
\citep{minezaki2009,macleod2013}. Note that our model 
is consistent with the best-fittingted macro model in \citet{macleod2013}
without a possible subhalo G3.
\subsection{B1608+656}
The fold caustic lens B1608+656 consists of three bright images A, B, C
and one faint image D. A source quasar at $z_S=1.394$ is lensed by two
early-type galaxies G1 and G2 at $z_L=0.630$
\citep{myers1995,fassnacht1996}. The measured ellipticity and the 
position angle of near-IR light distribution are $e=0.45\pm 0.01$ and 
$\theta_e=73.50\pm0.40(^\circ)$ for G1 and $e=0.55\pm 0.01$ and 
$\theta_e=-81.10\pm0.20(^\circ)$ for G2 \citep{sluse2012}.
The lens galaxy G1 belongs to a low-mass group of eight members
\citep{fassnacht2006}. We use the astrometry of lensed images and the centroid of 
G1 and G2 in \citet{sluse2012}. The fluxes and time delays between these four
images are based on radio-wavelength monitoring with the Very Large Array at 8.5 GHz \citep{fassnacht2002}. 
Time delay between image A and B is denoted as
$t_{\tr{A}}-t_{\tr{B}}=\Delta t_{\tr{BA}}$.
All the observed data are fitted well
by an SIE(for G1)+ES(for environment)+SIE(for G2) model
$\chi^2_{\tr{tot}}/\textrm{dof}=1.6$ though the best-fitting ellipticities $e(\tr{G1})=0.621,
e(\tr{G2})=0.759$ are somewhat larger than the observed values 
in \citet{sluse2012}. The
best-fitting Hubble constant $h=0.905$ is too large. However, $\sim 20\%$
deviation could be explained by deviation from the assumed power law 
of mass distribution \citep{schneider2013}. We
conclude that this system is not anomalous in the flux ratios though the
cusp relation is violated as $R_{\tr{cusp}}\sim 0.492$. This is due to
the complex structure of the lens.
\subsection{B0712+472}
The fold-caustic lens B0712+472 consists of two bright images A and B, and 
two fainter images C and D. The source and lens redshifts are
$z_S=1.339$ and $z_L=0.4060$ \citep{fassnacht1998}. We use the radio 
flux ratios in \citet{koopmans2003} of four images at 5 GHz
averaged over a period of 8.5 months. For the astrometry, 
we use the data from CASTLES data base. 
The positions of
lensed images and the centroid G of the primary lensing galaxy are well
fitted by an SIE and an ES assuming that the error in the
position of G is 0.05 arcsec. However, the flux ratios are not 
well fitted. For computing 
the magnification perturbation, we use only three bright images
as the signal-to-noise ratio of D/A is significantly worse than
the other images. 
\subsection{B2045+265}
The cusp-caustic lens B2045+265 consists of three bright images A, B, 
and C and one faint image D. The source and lens redshifts are
$z_S=1.28$ and $z_L=0.8673$ \citep{fassnacht1999}. We use the radio 
flux ratios \citep{koopmans2003} of four images at 5 GHz
averaged over a period of 8.5 months. For the astrometry, 
we use the infrared components of B2045 obtained by adaptive optics
imaging at $2.2 \mu \textrm{m}$\citep{mckean2007}. In addition to
a primary lensing galaxy G1 at $z_L$, a possible companion galaxy G2
may reside near G1 though the redshift has not been known\citep{mckean2007}.
All the observed data are fitted extremely well
by an SIE(for G1)+ES(for environment)+SIE(for G2) model giving 
$\chi^2_{\tr{tot}}/\textrm{dof}=0.03$. However, the best-fitting ellipticity of G2 seems 
too large $e(\tr{G2})=0.867$. Such a large 
value can only be expected from either
an edge-on disc system or a tidally disrupted dwarf galaxy. 
If we do not include an SIE for G2 in the lens model, it has been known 
that B2045+265 exhibits
strong anomaly in the flux ratios between three cusp-caustic 
images \citep{keeton2003}. Since G2 resides at a position between 
G1 and the three cusp-caustic images A, B and C, it is natural 
to include the lensing effect for G2.
Thus, we conclude that this system is not
anomalous in the flux ratios though the cusp relation is significantly violated 
as $R_{\tr{cusp}}\sim 0.501$.

\begin{table*}
\hspace{-1.5cm}
\begin{minipage}{170mm}
\caption{Quadruple lens systems}
\setlength{\tabcolsep}{2pt}
\begin{tabular}{llcllll}
\hline
\hline
  Lens system  & Image(type) & Position(obs.)($''$) & Flux ratio(obs.) &
    $\mu$(model) & Flux ratio(model) & References \\ 
\hline
  B1422+231 & A(I) & $(-0.3860\pm0.0004,0.3169\pm0.0003)$ & A/B=$0.9416\pm0.0080$ &
   $ 6.892$ & A/B=$0.7882$ & (1) (2)\\ 
  $~z_L=0.34$ & B(II) & $(0.,0.)$ && $-8.744$ & \\
  $~z_S=3.62$ & C(I) & $(0.3360\pm0.0003,-0.7516\pm0.0005)$ & C/B=$0.5188 \pm 0.0079$ &
   $4.327$ & C/B=$0.5070$ \\
  $~\varepsilon=1.4\,\textrm{mas}$& D(II) &
	 $(-0.9470\pm0.0006,-0.8012\pm0.0005)$ & D/B=$0.0226\pm 0.0057$
	     & $-0.334$ & D/B=$0.0368$ \\
  & G & $(-0.7321\pm0.0037,-0.6390\pm0.0054)$  \\
\hline
  B0128+437 & A(I) & $(0.000\pm 0.002,0.0000 \pm 0.0003)$ && $5.005$ & &
			 (1) (3) (4) \\ 
  $~z_L=1.145(\star)$ & B(II) & $(-0.099\pm0.003,0.095\pm0.003)$ & B/A=$0.584 \pm 0.030$ & $-2.940$ & B/A=$0.587$ \\
  $~z_S=3.124$ & C(I) & $(-0.521\pm0.004,-0.170 \pm 0.002)$ & C/A=$0.520 \pm 0.030$ & $2.323$ & C/A=$0.464$ \\
  $~\varepsilon=4.2\,\textrm{mas}$ & D(II) & $(-0.109\pm0.003,-0.260\pm0.002)$ & D/A=$0.506\pm 0.032$ & $-2.613$ & D/A=$0.522$\\
  & G & $(-0.217\pm0.01,-0.104\pm0.01)$  \\
\hline
  MG0414+0534 & A1(I) & $(-0.600\pm0.003,-1.942\pm0.003)$ && $16.593$ &
		   &  (5) (6) (7) \\ 
  $~z_L=0.96$ & A2(II) & $(-0.732\pm0.003,-1.549\pm0.003)$ &
	     A2/A1=$0.919 \pm 0.021$ &$-17.233$ & A2/A1=$1.007$ \\
  $~z_S=2.639$ & B(I) & $(0.,0.)$ & B/A1=$0.347 \pm 0.013$ & $5.456$ & B/A1=$0.341$ \\
  $~\varepsilon=4.2\,\textrm{mas}$ & C(II) &
	 $(1.342\pm0.003,-1.650\pm0.003)$ & C/A1=$0.139\pm 0.014$ &
		 $-2.704$ & C/A1=0.171 \\
  & G & $(0.472\pm0.003,-1.277\pm0.003)$ &&& \\
  & X & $(0.857\pm0.011,0.180\pm0.009)$ &&& \\
\hline
 B1608+656 & A(I) & $(0.,0.)$ & A$=3.41\pm 0.07$ & $4.904$ & A$=3.32$
		      & (2) (8) \\ 
  $~z_L=0.630$ & B(I) & $(-0.7464\pm0.0026,-1.9578\pm0.0026)$
	 &B$=1.68\pm 0.03$& $2.518$ & B$=1.70$  
   \\
  $~z_S=1.394$ & C(II) & $(-0.7483\pm 0.0038,-0.4465\pm 0.0033)$
	 &C$=1.73\pm 0.03$ &
	     $-2.576$ & C$=1.74$   \\
   $~\varepsilon=2.4\,\textrm{mas}$& D(II) &
	 $(1.1181\pm0.0025,-1.2527\pm0.0018)$ & D$=0.59\pm 0.01$  & $-0.8692$ & D$=0.588$   \\
  & G1 & $(0.4561\pm0.0061,-1.0647\pm0.0037)$ \\
  & G2 & $(-0.2821\pm0.0015,-0.9359\pm0.0023)$  \\
$(\Delta t_{\textrm{BA}}, \Delta t_{\textrm{BC}},\Delta
     t_{\textrm{BD}})$(days)& &
     $(31.5^{+2}_{-1},36.0\pm 1.5, 77.0^{+2}_{-1})$ & & & &  \\
\hline
  B0712+472 
& A(I) & $(0.795\pm0.003,-0.156\pm0.003)$ & & $8.716$ & & (1) (5)  \\ 
  $~z_L=0.406$ & B(II) & $(0.747\pm0.003,-0.292\pm0.006)$
	 &B/A$=0.843\pm 0.061$& $-7.735$ & B/A$=0.888$  
   \\
  $~z_S=1.339$ & C(I) & $(-0.013\pm 0.004,-0.804\pm 0.003)$
	 &C/A$=0.418\pm 0.037$ &
	     $3.051$ & C/A$=0.350$   \\
   $~\varepsilon=6.4\,\textrm{mas}$& D(II) &
	 $(-0.391\pm0.006,-0.082\pm0.003)$ & D/A$=0.082\pm 0.035$  & $-0.504$ & D/A$=0.0579$   \\
  & G & $(0.,0.)$ \\

\hline  
B2045+265 & A(I) & $(0.0000\pm0.0005,0.0000\pm0.0005)$ && $9.515$ & & (1)
		     (9) \\ 
  $~z_L=0.8673$ & B(II) & $(0.1316\pm0.0006,-0.2448\pm0.0006)$ &B/A$=0.578 \pm 0.059$ & $-5.531$ &
    B/A$=0.581$  \\
  $~z_S=1.28$ & C(I) & $(0.2869\pm 0.0005,-0.7885\pm 0.0005)$ & C/A$=0.739 \pm 0.073$ & $7.148$ & C/A$=0.751$  \\
  $~\varepsilon=1.4\,\textrm{mas}$&  D(II) & $(-1.6268\pm0.0013,-1.0064\pm0.0013)$ &D/A$=0.102\pm 0.025$ & $-0.970$ & D/A$=0.102$ \\
&   G1 & $(-1.1084\pm0.0011,-0.8065\pm0.0011)$ &&& \\
  & G2 & $(-0.4498\pm0.0021,-0.6425\pm0.0021)$ &&& \\
\hline
\label{table1}
\end{tabular}
 Note: ($\star$): The lens redshift $z_L$ is obtained from a best-fitting
 model. References: (1) \citet{koopmans2003} (2) \citet{sluse2012}  (3) \citet{biggs2004} (4) \citet{lagattuta2010}
(5) CASTLES data base:http://www.cfa.harvard.edu/castles
 (6) \citet{minezaki2009}    (7) \citet{macleod2013} (8)
 \citet{fassnacht2002} (9)\citet{mckean2007} Types I and II correspond
 to minimum and saddle, respectively. $\mu$ represents magnification. 
\end{minipage}
\end{table*}

\begin{table*}
\hspace{-4.1cm}
\begin{minipage}{138mm}
\caption{best-fitting model parameters}
\label{symbols}
\begin{tabular}{@{}lcccccc}
\hline
\hline
Model & B1422+231 &B0128+437 & MG0414+0534 & B1608+656 & B0712+472
 &B2045+265 \\
 & SIE-ES+($0.01''$) &SIE-ES& SIE-ES-SIS &SIE-ES-SIE& SIE-ES+($0.05''$) &
			 SIE-ES-SIE \\
\hline
 $b_{G1}'('')$ & 0.754 & 0.207& 1.07 &0.737  &0.543 & 1.012 \\
\hline
$(x_s,y_s)('')$ &  (-0.3854,-0.4144) &(-0.2549, -0.1001)& (0.4037,-1.0268) &
	(0.1027,-1.0981) & (0.0184,-0.1503) & (-0.6387,-0.6533) \\
\hline
$e(\tr{G1})$ & 0.300 & 0.577 & 0.300 &0.621  & 0.735 & 0.358\\
\hline
$\theta_{e(\tr{G1})}$(deg)&  -56.6 & -20.2 & -87.9 &73.1 & 57.2& 22.0 \\
\hline
$\gamma$ & 0.168 & 0.230 & 0.0870 &0.135 &0.199 & 0.220  \\
\hline
$\theta_\gamma$(deg) & -52.4 & 46.3 & 47.4 & -84.3 &-23.7 & -70.1 \\
\hline
$b_\textrm{G2}'('')$ & & &0.192 &0.212 & &0.0449 \\
\hline
$e(\tr{G2})$ & & & & 0.759 & &0.867 \\
\hline
$\theta_{e(\tr{G2})}$(deg) & & & & 63.6 & &-58.8 \\
\hline
 $(x_{\tr{G2}},y_{\tr{G2}})('')$ & & &(0.856,0.183) &  (-0.282, -0.936)  
& & (-0.450,-0.642)  \\
\hline
$(\delta x_{\tr{G1}}, \delta y_{\tr{G1}})('')$ & (-0.002, -0.016) &(0.013,-0.005)
 &(0.001,-0.003) &(-0.0001,0.0022) & (-0.102, 0.059 ) & (0.0000,0.0000)  \\
\hline
$\Delta t_{\textrm{BA}}$(days) & & & &32.9  & & \\
\hline
$\Delta t_{\textrm{BC}}$(days) & & & &37.1 & & \\
\hline
$\Delta t_{\textrm{BD}}$(days) & & & &75.8 & & \\
\hline
$h$(Hubble constant) & & & &0.905 & & \\
\hline

dof & 4& 4& 3 &3 & 4& 1\\
\hline
$\chi^2_{\tr{imag}}$  &1.4 & 2.1 &5.1 & 0.1& 4.2 & 0.00 \\
\hline
$\chi^2_{\tr{flux}}$  &340.3 &3.7 &22.9 &2.5 &4.4 &0.03 \\
\hline
$\chi^2_{\tr{tdel}}$  &  &  &  &2.0  & & \\
\hline
$\chi^2_{\tr{tot}}$  &344.1 &7.7 &29.2 & 4.8 & 14.1 &0.03 \\
\hline
\end{tabular}
\medskip
\end{minipage}
\end{table*}

\section{Results}

As shown in Table \ref{table3}, the observed magnification
perturbations $\hat{\eta}$ with respect to the best-fitting 
lens models in Section 4 are in the range of $0.063<\hat{\eta}<0.13$.
Using three lensed images ($N_{\tr{image}}=3$), we find that
$\eta$ for B1422+231 is non-zero at $\sim 20 \sigma$ level,
implying that the flux-ratio anomaly is most prominent.  
For the other three lensing systems, 
the significance of non-zero $\eta$ is $2\sim 3 \sigma$.

\begin{figure}
\hspace{-0.7cm}
\IG[width=92mm]{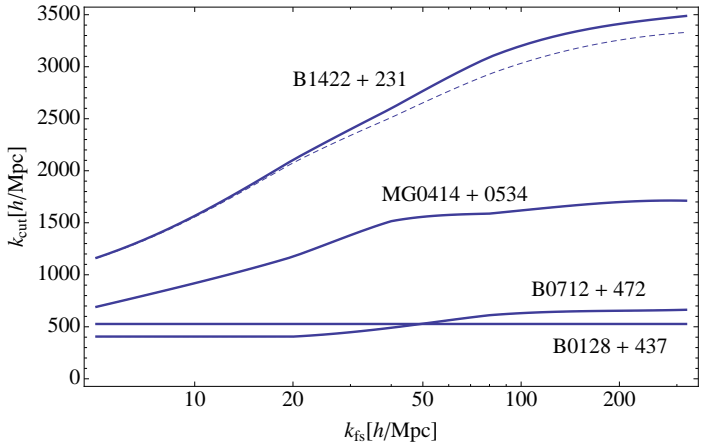}
\caption{Dependence of $k_{\tr{cut}}$ on the free streaming scale. $k_{\tr{cut}}$
is plotted as a function of $k_{\textrm{fs}}$ for $k_{\textrm{max}}=10^5\, 
h/\textrm{Mpc}$(full curve) and $k_{\textrm{max}}=3\times 10^3\, 
h/\textrm{Mpc}$(dashed curve). Except for B1422+231, the dashed curves 
almost coincide with the full curves. }
\label{f5}
\end{figure}

In order to estimate the second moment of $\eta$, we have to 
consider a cut-off scale $k_{\tr{max}}$ that corresponds to the smallest
fluctuations due to the finite size of the source. From dust
reverberation, the radius of the MIR emitting region of MG0414+0534 
is estimated as $\sim 2\,$pc\citep{minezaki2009}. As the magnifications of A1 and A2 images
are $\sim 17$, the apparent comoving size of the lensed source at the lens plane
is $r_s \sim (1+z_s)\times 2\times \sqrt{17}=30\,$pc. Assuming that 
$k_{\tr{max}}=2 \pi/(4 r_s)$, we have $k_{\tr{max}}\sim 8\times 10^4\, h/\tr{Mpc}$.   
For radio sources, we can estimate $k_{\tr{max}}$ from
the apparent angular sizes (typically $1\sim 3\,$mas in radius) of
lensed very long baseline interferometry (VLBI) images.
Then we find that $3\times 10^3 \lesssim k_{\tr{max}} \lesssim 1\times 10^5$ in units
of $h/\tr{Mpc}$. Taking into account ambiguity in the source size, we
consider two types of choices $k_{\tr{max}}=3\times 10^3\,h/\tr{Mpc}$
and $10^5\,h/\tr{Mpc}$. The corresponding $k_{\tr{cut}}$'s for the
anomalous four lenses are in the range
of $530< k_{\tr{cut}} <3560 \,h/\tr{Mpc}$ (Fig. \ref{f5}). We find that
$k_{\tr{cut}}$'s for the WDM models are equal to or less than the values
for the CDM model. A larger free-streaming scale yields a larger cut-off
scale $\sim 1/k_{\tr{cut}}$. Dependence of $k_{\tr{cut}}$ on 
$k_{\tr{max}}$ is found to be very small. The expected rms $\eta$'s 
for B1608+656 and B2045+265 are consistent with the null
result (Table 5).

\begin{figure}
\hspace{-0.7cm}
\IG[width=92mm]{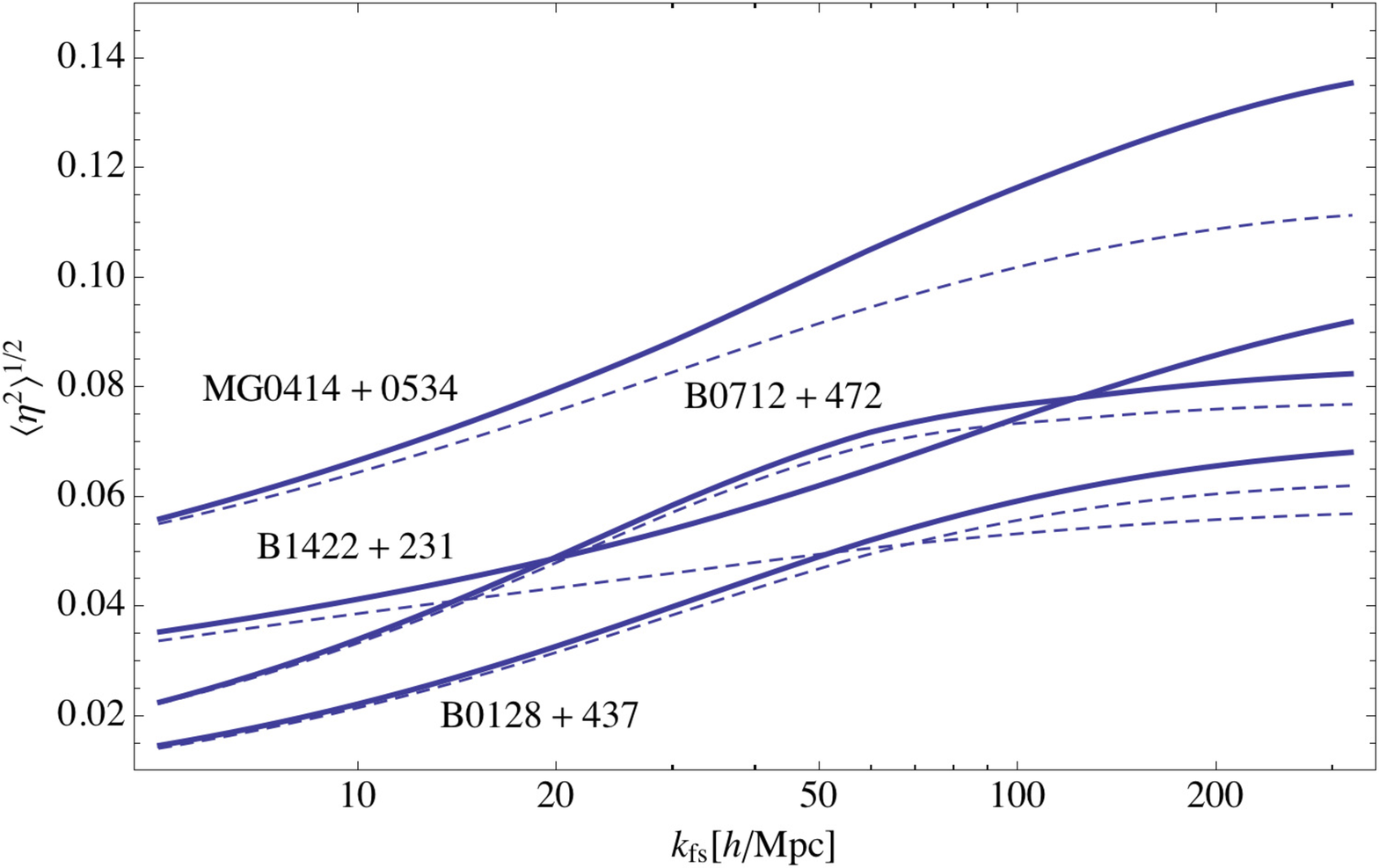}
\caption{Suppression of magnification perturbation due to free-streaming. The square-root of 
second moment $\langle \eta^2 \rangle^{1/2} $ 
is plotted as a function of $k_{\textrm{fs}}$ for $k_{\textrm{max}}=10^5\, 
h/\textrm{Mpc}$(full curve) and $k_{\textrm{max}}=3\times 10^3\, 
h/\textrm{Mpc}$(dashed curve).}
\label{plots-eta}
\end{figure}

\begin{figure}

\hspace{-0.5cm}
\IG[width=90mm]{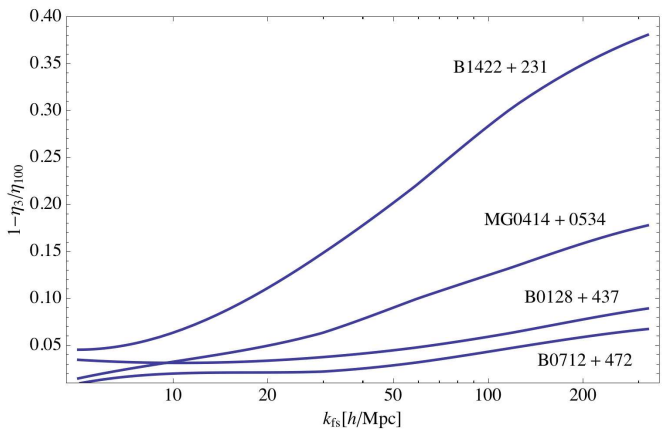}
\caption{Effect of small-scale fluctuations. 
$\eta_3$ and $\eta_{100}$ correspond to  $\langle \eta^2 \rangle^{1/2}$ for
 $k_{\textrm{max}}=3\times 10^3\, h/\textrm{Mpc}$ and
 $k_{\textrm{max}}=10^5\, h/\textrm{Mpc}$, respectively. 
}
\label{plots-of-eta3-eta100N}
\end{figure}

In the CDM model, we find that contribution 
from modes with wavelength $k>3\times 10^3\,h/\tr{Mpc}$ is not negligible, 
especially for high-redshift sources (Table 5). This suggests that 
$\eta$ in CDM models is sensitive on the property of small-scale fluctuations 
in systems with a high-redshift source. However, for the WDM models with
a large free-streaming scale, contribution 
from modes on small scales is very small (Fig. \ref{plots-eta}). For instance, 
the difference in the second moment of $\eta$ between the model with
$k_{\tr{max}}=3\times 10^3\,h/\tr{Mpc}$ and that with
$k_{\tr{max}}=10^5\,h/\tr{Mpc}$ is less than 15 $\%$ for
$k_{\tr{fs}}=30\,h/\tr{Mpc}$ (Fig. \ref{plots-of-eta3-eta100N}). 
The reason is as follows. As the squared amplitudes of convergence perturbation is 
proportional to $k^2 P(k)$ regardless of the free-streaming scale
(see Fig. 3), the modes with $k \sim k_{\tr{lens}}$ contribute much to the 
magnification perturbation $\eta$. Moreover, the decay of power due to
the free-streaming of WDM particles further reduces the contributions from small-scale
modes with $k> k_{\tr{lens}}$. 

\begin{figure}
\hspace{-0.66cm}
\IG[width=85mm]{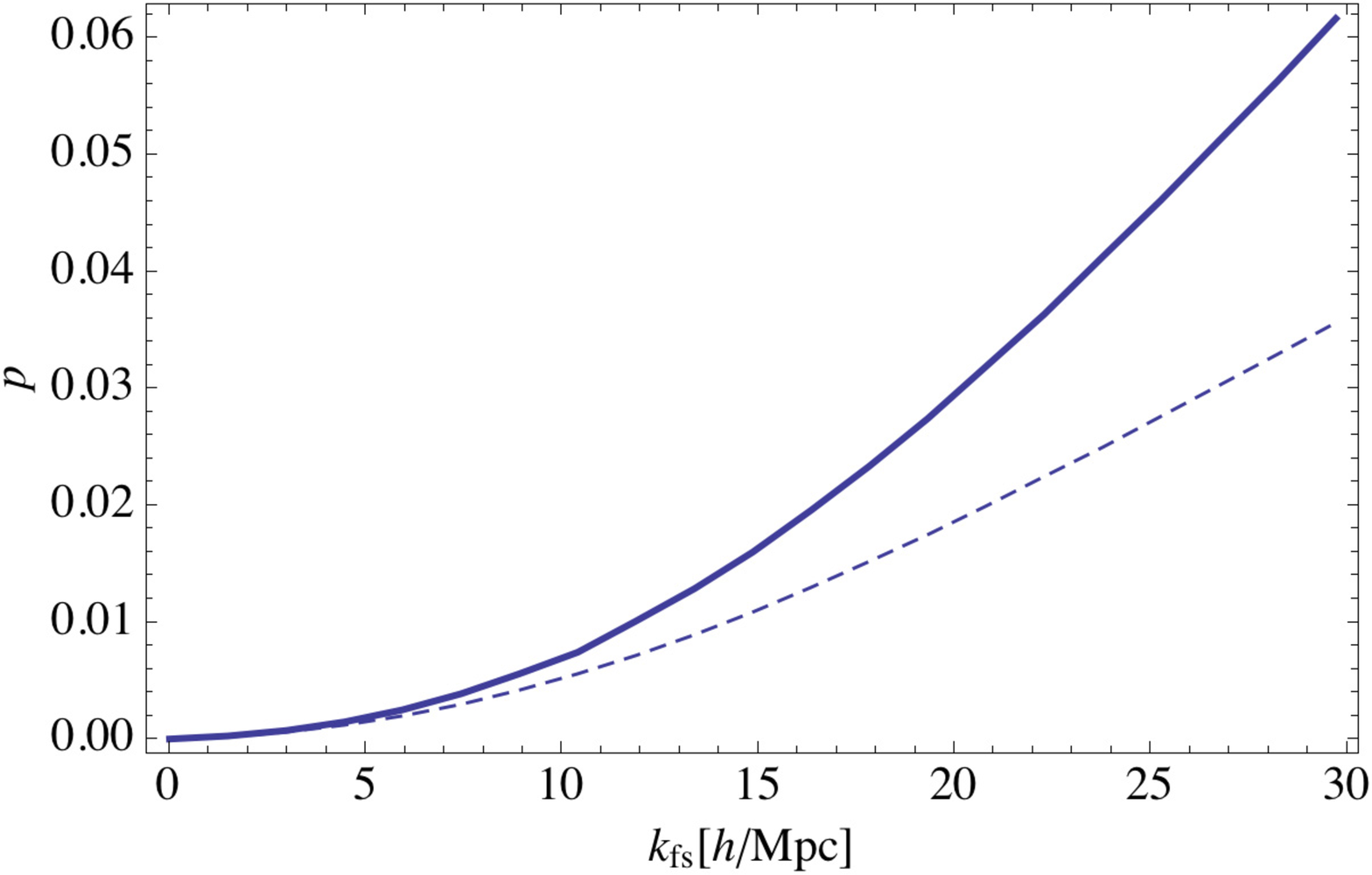}
\caption{Plots of $p$-value   
as a function of $k_{\textrm{fs}}$ for
$k_{\textrm{max}}=10^5\, 
h/\textrm{Mpc}$(full curve) and $k_{\textrm{max}}=3\times 10^3\, 
h/\textrm{Mpc}$(dashed curve).}
\label{fig:constraints-on-kf}
\end{figure}

In order to constrain the mass of WDM particles with a
free-streaming scale $k_{\textrm{fs}}$, we use the
PDFs of magnification perturbation $\eta$ for each 
anomalous lens system $i$. 
The PDF for system $i$ is $P(\eta_i;\langle \eta_i^2
\rangle^{1/2};\delta \eta_i )$
where $\langle \eta_i^2 \rangle$ is the second moment of $\eta_i$, which
is a function of $k_{\textrm{fs}}$ and $\delta \eta_i$ is the 1-$\sigma$ observational 
error for a lens $i$ (see Appendix B). For the null hypothesis that 
the observed non-vanishing $\eta_i$'s are due to line-of-sight structures 
in the WDM model, for $N$ anomalous systems with observed magnification
perturbations $(\hat{\eta}_1,\hat{\eta}_2, \cdots,\hat{\eta}_N)$, 
the $p$-value can be estimated as
\BE
p(k_{\textrm{fs}})=\frac{\displaystyle \int_S d \v{\eta} \prod_{i} P(\eta_i;\langle \eta_i^2 \rangle^{1/2};\delta \eta_i )}{\displaystyle\int_{all} d \v{\eta} \prod_{i} P(\eta_i;\langle \eta_i^2 \rangle^{1/2};\delta \eta_i )},
\EE
where $\v{\eta}=(\eta_1,\eta_2, \cdots , \eta_N)$ and a domain $S$ is
defined as a region where
\BE
\prod_{i} P(\eta_i;\langle \eta_i^2
\rangle^{1/2};\delta \eta_i ) < \prod_{i} P(\hat{\eta}_i;\langle \eta_i^2
\rangle^{1/2};\delta \eta_i )
\EE
holds. For the concordant $\Lambda$CDM model, we find that 
$p=0.19$ for $k_{\textrm{max}}=3\times 10^3\, h/\textrm{Mpc}$
and $p=0.53$ for $k_{\textrm{max}}=10^5\, h/\textrm{Mpc}$. Thus,
the null hypothesis cannot be rejected. For WDM models,
we find that $p<0.05$ if $k_{\textrm{fs}}<27 \, h/\textrm{Mpc}$
provided that $k_{\textrm{max}}=10^5\, h/\textrm{Mpc}$ (Fig. 
\ref{fig:constraints-on-kf}). In terms of the thermal WDM mass,
the constraint can be expressed as $m_{\textrm{WDM}}\ge
1.3\,\textrm{keV}$. For the mass of sterile neutrinos, 
the constraint corresponds to $m_{\textrm{s}}\gtrsim
 6.3\,\textrm{keV}$.
For smaller $k_{\textrm{max}}$, the constraint becomes more stringent.
Because the power spectrum on small scales obtained 
from our fitting function is systematically larger than
the correct values, the obtained constraint on the WDM mass 
is a conservative one.

\begin{table*}
\hspace{-0.7cm}
\begin{minipage}{170mm}
\caption{Magnification perturbation in CDM models}
\setlength{\tabcolsep}{2pt}
\begin{tabular}{lcccccccccc}
\hline
\hline
Lens system &$z_S$ & $z_L$ & $k_{\textrm{lens}}(h/\tr{Mpc})$
 &$k_{\textrm{source}}(h/\tr{Mpc})$ & $N_{\textrm{image}}$ & 
$\hat{\eta}$ & $\langle \eta^2 \rangle^{1/2 (\tr{CDM})}_{k_{\textrm{max}}=3\times
 10^3}(*)$ &$\langle \eta^2 \rangle^{1/2(\tr{CDM})}_{k_{\textrm{max}}=10^5}(*)$  
\\
\hline
B1422+231 &3.62 &0.34  &412 & $\sim 7\times 10^4$ &3 &$0.098\pm0.005$  &
			     0.058(3330) &0.10(3560)
\\
\hline 
B0128+437 &3.124 &1.145 &527 &$\sim 1 \times 10^4$  &4 &$0.0632\pm
			 0.025$ & 0.063(530) & 0.083(530)

\\
\hline
MG0414+0534 &2.639 &0.96 &118 &$\sim 8\times 10^4$ &4 &$0.131\pm
			 0.042$ & 0.11(1670) & 0.14(1720)
\\
\hline
B1608+656 &1.394 &0.630 &172 &$\gtrsim 3 \times 10^3$ &4 &$0.0223\pm
			 0.0091$ & 0.019(2240) & 0.026(2290)

\\
\hline 
B0712+472 &1.339 &0.406 &401 &$\sim 7 \times 10^4$ & 3 & $0.131\pm
			 0.071$ & 0.078(670) & 0.084(670)
\\
\hline
B2045+265 &1.28 &0.8673 &134 &$\sim 8 \times 10^4$ &3 &$0.075\pm
			 0.050$ & No solution & 0.049(3580)
\\
\hline
\label{table3}
\end{tabular}
\\
(*): The values inside parentheses indicate the corresponding
 $k_{\tr{cut}}$ in units of $h/\tr{Mpc}$. We estimate the 
$1\sigma$ errors in $\hat{\eta}$ by assuming that errors in the observed 
fluxes (flux ratios) obey the Gaussian statistics with no correlation
between the errors. 

\end{minipage}
\end{table*}

\section{Conclusion and Discussion}
We have investigated the weak lensing effect by line-of-sight structures
in a concordant CDM and WDM models based on $N$-body simulations. 
We have found that four quadruple lenses with source 
redshifts at $1 \le z_s \le 4$ out of
six show anomalies in the flux ratios of lensed images assuming that the
density of the primary lens is described by an SIE. 
The magnitudes of expected magnification perturbation due to the
line-of-sight structures
in the concordant $\Lambda$CDM model are consistent with the observed ones.
Using four anomalous samples and extrapolated power spectra obtained
from numerical simulations of WDM models, a constraint on the free-streaming
scale of WDM particles, $k_{\tr{fs}}\ge
27\,h/\tr{Mpc}(95\%\tr{CL})$ has been obtained. For thermally
produced WDMs, we have a constraint
$m_{\tr{WDM}}\ge 1.3 \,\tr{keV}(95\%\tr{CL})$.

Our result for fluctuations at low-redshifts $0<z<4$ is consistent 
with constraints from Lyman-$\alpha$ forests at
$4<z<6$ \citep{viel2013} and those 
from high-redshift ($4\lesssim z\lesssim 10$) galaxy counts
\citep{schultz2014}. Therefore, WDM models with $m_{\tr{WDM}}< 1.3
\,\tr{keV}$ 
are ruled out at
redshifts $0.5\lesssim  z \lesssim 10$. Thus, WDM models as solutions for the
'missing satellite problem' are disfavoured virtually at all the redshifts.
 
Our calculations are based on a semi-analytic formalism that has been
used for estimating magnification perturbations due to line-of-sight
structures in the CDM models. In order to verify 
the assumed PDF form of magnification
perturbation $\eta$ in WDM models (see Appendix B), 
we need ray-tracing Monte Carlo simulations where
the lens parameter fitting is done with the presence of line-of-sight 
structures, which will be our future work.

In our simulations, we did not take into account non-luminous subhaloes in
lensing galaxies. In CDM models, it has been shown that 
the surface mass density of subhaloes in lensing galaxies are not enough
for explaining the observed flux-ratio anomalies \citep{amara2006,
maccio2006,chen2009,xu2009,xu2010,chen2011}. 
As the number density of subhaloes with sizes that are comparable to or less than the
free-streaming scale $\sim 1/k_{\tr{fs}}$ is significantly reduced, 
the role of dark subhaloes in lensing galaxies would be minor. However,  
we may need to check the lensing effects of subhaloes in WDM models as well.  

We expect that baryonic feedback effects on the line-of-sight
structures are limited to the central region of
minihaloes. Therefore, the weak lensing effects that are relevant to the
property of outskirts of minihaloes may not change so much. Although, the 
power spectra from simulations with baryons may significantly differ, those
calculated from fluctuations obtained by masking the central regions of haloes
would be less affected. In order to verify it, however,  
it is very important to incorporate baryonic effects in our lensing
simulations, which will also be our future work.

Another important issue is the ambiguity in the macro lens model.
Although we have assumed that the density of the macro lens can
be described by an SIE plus an ES, the actual lens may 
have a more complex structure, such as anisotropy in the velocity
dispersion and deviation from power laws, etc. Therefore, 
caution has to be made that the constraints on the WDM mass could be 
weakened if the applied macroscopic mass models based on SIEs with
a constant ES are not proper ones \citep{xu2014}. 

However, in the near future, we will 
obtain a larger sample of lens and more precise
information about the macro lens model [e.g., the Atacama Large
Millimeter/submillimeter Array (ALMA) and Thirty Meter Telescope],
which have a potential for breaking degeneracy in the lens model.    
 
If intervening perturbers are massive enough ($\gtrsim 10^{10} \ms$),
we may directly detect the presence and the redshift of
perturbers from the extended-source effects \citep{inoue2005b,
inoue2005a}. In order to do so, observation of anomalous quadruple lenses by ALMA is important.
Emission from neutral hydrogen (HI) may be another
clue for detecting the line-of-sight structures. Measuring correlation between
flux-ratio anomaly and HI emission may be a new test for confirming the
presence of line-of-sight structures.  
\section{Acknowledgements}
We thank the anonymous referees for his/her valuable comments. 
This work is supported in part by JSPS Grant-in-Aid for
Scientific Research (B) (No. 25287062) ``Probing the origin
of primordial minihaloes via gravitational lensing phenomena''.
The work of TT is partially supported by the Grant-in-Aid for Scientific
Research from the Ministry of Education, Science, Sports, and Culture,
Japan, No.~23740195. TI is financially supported by MEXT HPCI 
Strategic Program and MEXT/JSPS KAKENHI Grant Number 24740115.
Numerical computations were carried out on Cray XT4 at 
Center for Computational Astrophysics, CfCA, at National 
Astronomical Observatory of Japan and the K computer at the
RIKEN Advanced Institute for Computational Science (proposal numbers
hp120286 and hp130026).
\appendix

\section{Fitting formula for non-linear matter power spectra in WDM models}

In this appendix, we present our fitting formula for the matter power
 spectra in WDM models.
Our formula is based on the halofit model \citep{smith2003,takahashi2012},
 but slightly modified for WDM models.

To find the best fitting parameters in the theoretical model, we use the
standard chi-square fitting, which is defined as
\begin{equation}
 \chi^2 = \sum_{i} \sum_{z=0}^{3} \sum_{k=k_{\rm min}}^{k_{\rm max}}
 \frac{\left[ P_{i, {\rm model}}(k,z)-P_{i, {\rm sim}}(k,z) \right]^2}
 {P_{i, {\rm sim}}(k,z)^2},
\end{equation}
where $P_{i, {\rm model}}$ is the power in the theoretical model, $P_{i, {\rm sim}}$'s are
 those in simulation results, and $i$ denotes 
the CDM model ($i=0$) and
 the four WDM models ($i=1-4$) in Table \ref{table_cdm_wdm}.
The $\chi^2$ is summed over redshifts $z=0, 0.3, 0.6, 1, 2$ and $3$.
We use the wavenumber $k$ larger than $2\,h /{\rm Mpc} (=k_{\rm min})$
 where the Gaussian error of $P(k)$ is less than $1\%$.
The maximum wavenumber is $k_{\rm max}=300 (30) h /{\rm Mpc}$
 for the L10(L100) so that the measured power spectrum is
 much larger ($10$ times larger) than the shot noise.

First, we fit the simulation results of the CDM model.
The fitting parameters for the CDM model are the same as in \citet{takahashi2012},
 except for the following three parameters:
\begin{eqnarray}
 && \hspace{-0.5cm} \log_{10} a_{\rm n} = 0.9221 + 2.0595 n_{\rm eff}
  + 2.4447 n_{\rm eff}^2+ 1.2625 n_{\rm eff}^3 \nonumber \\
 && \hspace{1.cm} + 0.2874 n_{\rm eff}^4 - 0.7601 C, \nonumber \\
 && \hspace{-0.5cm} \log_{10} c_{\rm n} = 0.4747 + 2.1542 n_{\rm eff}
 + 0.8582 n_{\rm eff}^2 + 0.8329 C,
 \nonumber  \\
 && \hspace{-0.5cm} \gamma_{\rm n} = 0.2247 - 0.2287 n_{\rm eff}
 + 0.9726 C - 0.0533 \ln \left( \frac{k}{h /{\rm Mpc}} \right).
 \nonumber \\
\label{halofit_params}
\end{eqnarray}
The ratios of the power spectra of the WDM to that of the CDM are fitted as,
\BE
 \frac{P_{\rm wdm}(k,z)}{P_{\rm cdm}(k,z)} = \frac{1}{\left( 1 +
 k/k_{\rm d} \right)^{0.7441}},
\label{pk_ratio}
\EE
with
\BE
 k_{\rm d} (k_{\rm fs},z) = 2.206 \, h {\rm Mpc}^{-1}
 \left( \frac{k_{\rm fs}}{h /{\rm Mpc}} \right)^{1.703}
 D(z)^{1.583},
\label{eq_kd}
\EE
where $k_{\rm fs}$ is the free-streaming wavenumber and $D(z)$ is the linear
 growth factor at $z$, which is normalized as $D(z=0)=1$.
Equation (\ref{eq_kd}) can be rewritten in terms of the WDM particle
 mass $m_{\rm WDM}$ as
\BE
 k_{\rm d} (k_{\rm fs},z) = 388.8 \, h {\rm Mpc}^{-1}
 \left( \frac{m_{\rm WDM}}{{\rm keV}} \right)^{2.027}
 D(z)^{1.583}.
\EE
The RHS of equation (\ref{pk_ratio}) corresponds to a damping factor.
Note that the parameters $n_{\rm eff}$ and $C$ in equation
 (\ref{halofit_params}) 
 are evaluated in the CDM model even when computing $P_{\rm wdm}(k,z)$.
Using our fitting formula, the simulation results can be reproduced
within a relative error of $19\%$. The rms deviation between the
 theoretical model and the simulation results is about $4.5 \%$.

\begin{figure}
%\epsscale{0.9}
\vspace*{1.0cm}
\includegraphics[width=85mm]{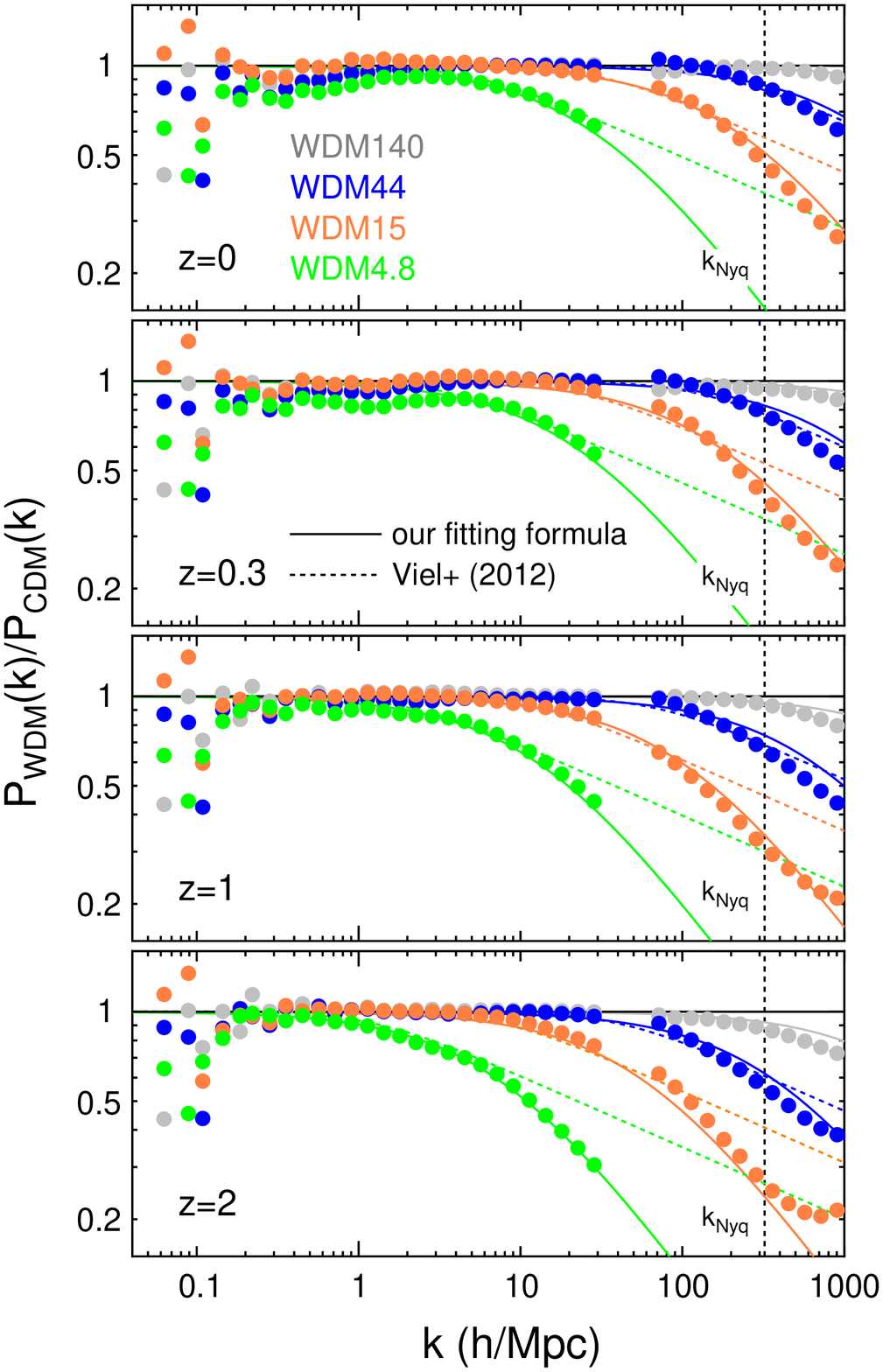}
\caption{
Same as Fig.\ref{fig_pk}, but the ratio of the WDM power spectra
 $P_{\rm WDM}(k)$ to the CDM $P_{\rm CDM}(k)$.
The filled circles are the simulation results for WDM140(grey),
 WDM44(blue), WDM15(orange) and WDM4.8(green).
The solid curves correspond to 
our fitting formula in equation (\ref{pk_ratio}), 
 while the dotted curves to the previous one in \citet{viel2012}.
}
\label{fig_pk_ratio}
\vspace*{0.5cm}
\end{figure}

Fig.\ref{fig_pk_ratio} shows the ratio of the WDM power spectrum
 $P_{\rm WDM}(k,z)$ to the CDM power spectrum $P_{\rm CDM}(k,z)$ at redshifts
 $z=0,0.3,1$ and $2$.
The solid curves represent the decay of power spectra described by  
our damping factor in equation (\ref{pk_ratio}), which
 reproduces our simulation results very well.
For comparison, the predicted power spectra based on the
previous fitting formula in \citet{viel2012} are plotted 
as dotted curves. As shown in Fig.\ref{fig_pk_ratio}, the 
predicted powers based on the previous fitting formula are too large
at scales $k \sim 100 h/$Mpc. The discrepancy becomes more
prominent at smaller scales $k \gtrsim 100 h/$Mpc.
This is probably due to the fact that 
the Nyquist wavenumber $k_{\tr{Nyq}}=60h/$Mpc (see fig.7 in \citet{viel2012}) 
in their simulations is smaller than ours $k_{\tr{Nyq}}=322h/$Mpc.

%We also comment that their formula can be used for $m_{\rm WDM} > 0.5$keV
% and thus it can not be used for our WDM4.8 model ($m_{\rm WDM}=0.29$keV).

\section{Functional Form of PDF}

In this appendix, we provide the PDFs of $\eta$. We assume that the PDFs 
are approximated by the log-normal function as
\BE
  P(\eta) = N \exp \left[ - \frac{1}{2 \sigma^2}
 \left\{ \ln \left( 1 + \frac{\eta}{\eta_0} \right) - \ln \mu \right\}^2
 \right] \frac{1}{\eta+\eta_0},
\label{pdf_eta}
\EE
where $N$ is a normalization constant, $\eta_0$ describes a dispersion
scale of $\eta$, and $\sigma$ and $\mu$ are constants. We assume that
$\eta_0$ depends only on the second moment $\langle \eta^2 \rangle $
and that $\sigma$ and $\mu$ do not depend on $\langle \eta^2 \rangle $.
Using ray-tracing simulations for a concordant $\Lambda$CDM model, we find 
that the best-fitting parameters are
\citep{takahashi-inoue2014}
\BE
 \mu=4.10, ~\sigma^2=0.279, ~\eta_0=0.228 \langle \eta^2 \rangle^{1/2}.
\label{pdf_eta_params}
\EE 
As the formula (\ref{pdf_eta}) does not depend on the grid size 
$r_{\textrm{grid}}$ of the simulation (see Fig.9 in \citet{takahashi-inoue2014}), we assume 
that the formula (\ref{pdf_eta}) is also applicable to WDM models
where small-scale fluctuations are suppressed due to
free-streaming\footnote{Even in 
CDM models, the slopes of density profiles of haloes on scales equal to or less
than the free streaming scale are different from 
those on larger scales (e.g. \citet{ishiyama2010, ishiyama2014}). 
However, the difference is striking only in the very
inner regions (within 10\% of the virial radius) of haloes. 
Since the contribution to the PDF of $\eta$ due
to the weak lensing effects mainly comes
from the outer regions ($>$virial radius), it would be 
reasonable to assume that the statistics of $\eta$ in WDM models are the same
as those in CDM models.}.
We find that the suppression in the power due to the finite grid size $\sim
4.8 \tr{kpc}/h$ is comparable to the one by free-streaming 
for $k_{\tr{fs}}\sim 20\, h/\tr{Mpc}$. 

In real setting, we need to take into account errors in observation as well.
If the variance of observational error is $\delta \eta^2$, and 
the mean value is vanishing, we should change $\eta_0$ as 
\BE
\eta_0=0.228 ( \langle \eta^2 \rangle+ \delta \eta^2)^{1/2}.
\EE 
             
\bibliographystyle{mn2e}
\bibliography{weak-lensing-by-los.bib}

\begin{thebibliography}{}

\bibitem[\protect\citeauthoryear{Amara, Metcalf, Cox \& Ostriker}{Amara
  et~al.}{2006}]{amara2006}
Amara A.,  Metcalf R.~B.,  Cox T.~J.,    Ostriker J.~P.,  2006, Monthly Notices
  of the Royal Astronomical Society, 367, 1367

\bibitem[\protect\citeauthoryear{{Angulo}, {Hahn} \& {Abel}}{{Angulo}
  et~al.}{2013}]{angulo2013}
{Angulo} R.~E.,  {Hahn} O.,    {Abel} T.,  2013, Monthly Notices of Royal
  Astronomical Society, 434, 3337

\bibitem[\protect\citeauthoryear{{Bagla} \& {Padmanabhan}}{{Bagla} \&
  {Padmanabhan}}{1997}]{bp1997}
{Bagla} J.~S.,  {Padmanabhan} T.,  1997, Monthly Notices of Royal Astronomical
  Society, 286, 1023

\bibitem[\protect\citeauthoryear{{Biggs}, {Browne}, {Jackson}, {York},
  {Norbury}, {McKean} \& {Phillips}}{{Biggs} et~al.}{2004}]{biggs2004}
{Biggs} A.~D.,  {Browne} I.~W.~A.,  {Jackson} N.~J.,  {York} T.,  {Norbury}
  M.~A.,  {McKean} J.~P.,    {Phillips} P.~M.,  2004, Monthly Notices of Royal
  Astronomical Society, 350, 949

\bibitem[\protect\citeauthoryear{{Bode}, {Ostriker} \& {Turok}}{{Bode}
  et~al.}{2001}]{bode2001}
{Bode} P.,  {Ostriker} J.~P.,    {Turok} N.,  2001, The Astrophysical Journal,
  556, 93

\bibitem[\protect\citeauthoryear{{Boehm}, {Mathis}, {Devriendt} \&
  {Silk}}{{Boehm} et~al.}{2005}]{boehm2005}
{Boehm} C.,  {Mathis} H.,  {Devriendt} J.,    {Silk} J.,  2005, Monthly Notices
  of Royal Astronomical Society, 360, 282

\bibitem[\protect\citeauthoryear{{Boyarsky}, {Lesgourgues}, {Ruchayskiy} \&
  {Viel}}{{Boyarsky} et~al.}{2009}]{boyarsky2009}
{Boyarsky} A.,  {Lesgourgues} J.,  {Ruchayskiy} O.,    {Viel} M.,  2009,
  Journal of Cosmology and Astroparticle Physics, 5, 12

\bibitem[\protect\citeauthoryear{Chen}{Chen}{2009}]{chen2009}
Chen J.,  2009, Astronomy \& Astrophysics, 498, 49

\bibitem[\protect\citeauthoryear{Chen, Koushiappas \& Zentner}{Chen
  et~al.}{2011}]{chen2011}
Chen J.,  Koushiappas S.~M.,    Zentner A.~R.,  2011, Astrophysical Journal,
  741

\bibitem[\protect\citeauthoryear{Chen, Kravtsov \& Keeton}{Chen
  et~al.}{2003}]{chen2003}
Chen J.,  Kravtsov A.~V.,    Keeton C.~R.,  2003, Astrophysical Journal, 592,
  24

\bibitem[\protect\citeauthoryear{Chiba}{Chiba}{2002}]{chiba2002}
Chiba M.,  2002, The Astrophysical Journal, 565, 17

\bibitem[\protect\citeauthoryear{Chiba, Minezaki, Kashikawa, Kataza \&
  Inoue}{Chiba et~al.}{2005}]{chiba2005}
Chiba M.,  Minezaki T.,  Kashikawa N.,  Kataza H.,    Inoue K.~T.,  2005,
  Astrophysical Journal, 627, 53

\bibitem[\protect\citeauthoryear{{Colombi}, {Dodelson} \& {Widrow}}{{Colombi}
  et~al.}{1996}]{Colombi1996}
{Colombi} S.,  {Dodelson} S.,    {Widrow} L.~M.,  1996, Astrophysical Journal,
  458, 1

\bibitem[\protect\citeauthoryear{Crocce, Pueblas \& Scoccimarro}{Crocce
  et~al.}{2006}]{crocce2006}
Crocce M.,  Pueblas S.,    Scoccimarro R.,  2006, Monthly Notices of the Royal
  Astronomical Society, 373, 369

\bibitem[\protect\citeauthoryear{{Dalal} \& {Kochanek}}{{Dalal} \&
  {Kochanek}}{2002}]{dalal-kochanek2002}
{Dalal} N.,  {Kochanek} C.~S.,  2002, The Astrophysical Journal, 572, 25

\bibitem[\protect\citeauthoryear{{Dodelson} \& {Widrow}}{{Dodelson} \&
  {Widrow}}{1994}]{Dodelson:1993je}
{Dodelson} S.,  {Widrow} L.~M.,  1994, Physical Review Letters, 72, 17

\bibitem[\protect\citeauthoryear{{Fassnacht}, {Blandford}, {Cohen}, {Matthews},
  {Pearson}, {Readhead}, {Womble}, {Myers}, {Browne}, {Jackson}, {Marlow},
  {Wilkinson}, {Koopmans}, {de Bruyn}, {Schilizzi}, {Bremer} \&
  {Miley}}{{Fassnacht} et~al.}{1999}]{fassnacht1999}
{Fassnacht} C.~D.,  {Blandford} R.~D.,  {Cohen} J.~G.,  {Matthews} K.,
  {Pearson} T.~J.,  {Readhead} A.~C.~S.,  {Womble} D.~S.,  {Myers} S.~T.,
  {Browne} I.~W.~A.,  {Jackson} N.~J.,  {Marlow} D.~R.,  {Wilkinson} P.~N.,
  {Koopmans} L.~V.~E.,  {de Bruyn} A.~G.,  {Schilizzi} R.~T.,  {Bremer} M.,
  {Miley} G.,  1999, Astrophysical Journal, 117, 658

\bibitem[\protect\citeauthoryear{{Fassnacht} \& {Cohen}}{{Fassnacht} \&
  {Cohen}}{1998}]{fassnacht1998}
{Fassnacht} C.~D.,  {Cohen} J.~G.,  1998, Astrophysical Journal, 115, 377

\bibitem[\protect\citeauthoryear{{Fassnacht}, {Gal}, {Lubin}, {McKean},
  {Squires} \& {Readhead}}{{Fassnacht} et~al.}{2006}]{fassnacht2006}
{Fassnacht} C.~D.,  {Gal} R.~R.,  {Lubin} L.~M.,  {McKean} J.~P.,  {Squires}
  G.~K.,    {Readhead} A.~C.~S.,  2006, The Astrophysical Journal, 642, 30

\bibitem[\protect\citeauthoryear{{Fassnacht}, {Womble}, {Neugebauer}, {Browne},
  {Readhead}, {Matthews} \& {Pearson}}{{Fassnacht}
  et~al.}{1996}]{fassnacht1996}
{Fassnacht} C.~D.,  {Womble} D.~S.,  {Neugebauer} G.,  {Browne} I.~W.~A.,
  {Readhead} A.~C.~S.,  {Matthews} K.,    {Pearson} T.~J.,  1996, The
  Astrophysical Journal, 460, L103

\bibitem[\protect\citeauthoryear{{Fassnacht}, {Xanthopoulos}, {Koopmans} \&
  {Rusin}}{{Fassnacht} et~al.}{2002}]{fassnacht2002}
{Fassnacht} C.~D.,  {Xanthopoulos} E.,  {Koopmans} L.~V.~E.,    {Rusin} D.,
  2002, The Astrophysical Journal, 581, 823

\bibitem[\protect\citeauthoryear{{Hewitt}, {Turner}, {Lawrence}, {Schneider} \&
  {Brody}}{{Hewitt} et~al.}{1992}]{hewitt1992}
{Hewitt} J.~N.,  {Turner} E.~L.,  {Lawrence} C.~R.,  {Schneider} D.~P.,
  {Brody} J.~P.,  1992, Astrophysical Journal, 104, 968

\bibitem[\protect\citeauthoryear{Hockney \& Eastwood}{Hockney \&
  Eastwood}{1988}]{hockney1988}
Hockney R.,  Eastwood J.,  1988, Computer simulation using particles.
Taylor \& Francis, New York

\bibitem[\protect\citeauthoryear{Inoue \& Chiba}{Inoue \&
  Chiba}{2005a}]{inoue2005a}
Inoue K.~T.,  Chiba M.,  2005a, Astrophysical Journal, 634, 77

\bibitem[\protect\citeauthoryear{Inoue \& Chiba}{Inoue \&
  Chiba}{2005b}]{inoue2005b}
Inoue K.~T.,  Chiba M.,  2005b, Astrophysical Journal, 633, 23

\bibitem[\protect\citeauthoryear{{Inoue} \& {Takahashi}}{{Inoue} \&
  {Takahashi}}{2012}]{inoue-takahashi2012}
{Inoue} K.~T.,  {Takahashi} R.,  2012, Monthly Notices of Royal Astronomical
  Society, 426, 2978

\bibitem[\protect\citeauthoryear{{Ishiyama}}{{Ishiyama}}{2014}]{ishiyama2014}
{Ishiyama} T.,  2014, Astrophysical Journal, 788, 27

\bibitem[\protect\citeauthoryear{{Ishiyama}, {Fukushige} \&
  {Makino}}{{Ishiyama} et~al.}{2009}]{ishiyama2009}
{Ishiyama} T.,  {Fukushige} T.,    {Makino} J.,  2009, Publications of the
  Astronomical Society of Japan, 61, 1319

\bibitem[\protect\citeauthoryear{{Ishiyama}, {Makino} \&
  {Ebisuzaki}}{{Ishiyama} et~al.}{2010}]{ishiyama2010}
{Ishiyama} T.,  {Makino} J.,    {Ebisuzaki} T.,  2010, Astrophysical Journal,
  723, L195

\bibitem[\protect\citeauthoryear{{Ishiyama}, {Nitadori} \& {Makino}}{{Ishiyama}
  et~al.}{2012}]{ishiyama2012}
{Ishiyama} T.,  {Nitadori} K.,    {Makino} J.,  2012, Proc. Int. Conf. High
  Performance Computing, Networking, Storage and Analysis, SC '12 (Los
  Alamitos, CA: IEEE Computer Society Press), 5, (arXiv:1211.4406)

\bibitem[\protect\citeauthoryear{{Jenkins}, {Frenk}, {Pearce}, {Thomas},
  {Colberg}, {White}, {Couchman}, {Peacock}, {Efstathiou} \&
  {Nelson}}{{Jenkins} et~al.}{1998}]{jenkins1998}
{Jenkins} A.,  {Frenk} C.~S.,  {Pearce} F.~R.,  {Thomas} P.~A.,  {Colberg}
  J.~M.,  {White} S.~D.~M.,  {Couchman} H.~M.~P.,  {Peacock} J.~A.,
  {Efstathiou} G.,    {Nelson} A.~H.,  1998, Astrophysical Journal, 499, 20

\bibitem[\protect\citeauthoryear{Keeton, Gaudi \& Petters}{Keeton
  et~al.}{2003}]{keeton2003}
Keeton C.~R.,  Gaudi B.~S.,    Petters A.~O.,  2003, Astrophysical Journal,
  598, 138

\bibitem[\protect\citeauthoryear{{Kolb} \& {Turner}}{{Kolb} \&
  {Turner}}{1990}]{kolb_turner}
{Kolb} E.~W.,  {Turner} M.~S.,  1990, {The early universe.}

\bibitem[\protect\citeauthoryear{{Koopmans}, {Biggs}, {Blandford}, {Browne},
  {Jackson}, {Mao}, {Wilkinson}, {de Bruyn} \& {Wambsganss}}{{Koopmans}
  et~al.}{2003}]{koopmans2003}
{Koopmans} L.~V.~E.,  {Biggs} A.,  {Blandford} R.~D.,  {Browne} I.~W.~A.,
  {Jackson} N.~J.,  {Mao} S.,  {Wilkinson} P.~N.,  {de Bruyn} A.~G.,
  {Wambsganss} J.,  2003, The Astrophysical Journal, 595, 712

\bibitem[\protect\citeauthoryear{Kormann, Schneider \& Bartelmann}{Kormann
  et~al.}{1994}]{kormann1994}
Kormann R.,  Schneider P.,    Bartelmann M.,  1994, Astronomy and Astrophysics,
  284, 285

\bibitem[\protect\citeauthoryear{Kundic, Hogg, Blandford, Cohen, Lubin \&
  Larkin}{Kundic et~al.}{1997}]{kundic1997b}
Kundic T.,  Hogg D.~W.,  Blandford R.~D.,  Cohen J.~G.,  Lubin L.~M.,    Larkin
  J.~E.,  1997, Astronomical Journal, 114, 2276

\bibitem[\protect\citeauthoryear{{Lagattuta}, {Auger} \&
  {Fassnacht}}{{Lagattuta} et~al.}{2010}]{lagattuta2010}
{Lagattuta} D.~J.,  {Auger} M.~W.,    {Fassnacht} C.~D.,  2010, The
  Astrophysical Journal, 716, L185

\bibitem[\protect\citeauthoryear{Lawrence, Elston, Januzzi \& Turner}{Lawrence
  et~al.}{1995}]{lawrence1995}
Lawrence C.~R.,  Elston R.,  Januzzi B.~T.,    Turner E.~L.,  1995,
  Astronomical Journal, 110, 2570

\bibitem[\protect\citeauthoryear{{Lewis}, {Challinor} \& {Lasenby}}{{Lewis}
  et~al.}{2000}]{camb}
{Lewis} A.,  {Challinor} A.,    {Lasenby} A.,  2000, Astrophysical Journal,
  538, 473

\bibitem[\protect\citeauthoryear{Maccio \& Miranda}{Maccio \&
  Miranda}{2006}]{maccio2006}
Maccio A.~V.,  Miranda M.,  2006, Monthly Notices of the Royal Astronomical
  Society, 368, 599

\bibitem[\protect\citeauthoryear{McKean, Koopmans, Flack, Fassnacht, Thompson,
  Matthews, Blandford, Readhead \& Soifer}{McKean et~al.}{2007}]{mckean2007}
McKean J.~P.,  Koopmans L. V.~E.,  Flack C.~E.,  Fassnacht C.~D.,  Thompson D.,
   Matthews K.,  Blandford R.~D.,  Readhead A. C.~S.,    Soifer B.~T.,  2007,
  Monthly Notices of the Royal Astronomical Society, 378, 109

\bibitem[\protect\citeauthoryear{{MacLeod}, {Jones}, {Agol} \&
  {Kochanek}}{{MacLeod} et~al.}{2013}]{macleod2013}
{MacLeod} C.~L.,  {Jones} R.,  {Agol} E.,    {Kochanek} C.~S.,  2013, The
  Astrophysical Journal, 773, 35

\bibitem[\protect\citeauthoryear{Mao \& Schneider}{Mao \&
  Schneider}{1998}]{mao1998}
Mao S.,  Schneider P.,  1998, Monthly Notices of the Royal Astronomical
  Society, 295, 587

\bibitem[\protect\citeauthoryear{{McKean}, {Koopmans}, {Browne}, {Fassnacht},
  {Blandford}, {Lubin} \& {Readhead}}{{McKean} et~al.}{2004}]{mckean2004}
{McKean} J.~P.,  {Koopmans} L.~V.~E.,  {Browne} I.~W.~A.,  {Fassnacht} C.~D.,
  {Blandford} R.~D.,  {Lubin} L.~M.,    {Readhead} A.~C.~S.,  2004, Monthly
  Notices of Royal Astronomical Society, 350, 167

\bibitem[\protect\citeauthoryear{Metcalf}{Metcalf}{2005}]{metcalf2005a}
Metcalf R.~B.,  2005, The Astrophysical Journal, 629, 673

\bibitem[\protect\citeauthoryear{Metcalf \& Madau}{Metcalf \&
  Madau}{2001}]{metcalf2001}
Metcalf R.~B.,  Madau P.,  2001, The Astrophysical Journal, 563, 9

\bibitem[\protect\citeauthoryear{Metcalf, Moustakas, Bunker \& Parry}{Metcalf
  et~al.}{2004}]{metcalf2004}
Metcalf R.~B.,  Moustakas L.~A.,  Bunker A.~J.,    Parry I.~R.,  2004,
  Astrophysical Journal, 607, 43

\bibitem[\protect\citeauthoryear{Minezaki, Chiba, Kashikawa, Inoue \&
  Kataza}{Minezaki et~al.}{2009}]{minezaki2009}
Minezaki T.,  Chiba M.,  Kashikawa N.,  Inoue K.~T.,    Kataza H.,  2009,
  Astrophysical Journal, 697, 610

\bibitem[\protect\citeauthoryear{Miranda \& Maccio}{Miranda \&
  Maccio}{2007}]{miranda2007}
Miranda M.,  Maccio A.~V.,  2007, Monthly Notices of the Royal Astronomical
  Society, 382, 1225

\bibitem[\protect\citeauthoryear{More, McKean, More, Porcas, Koopmans \&
  Garrett}{More et~al.}{2009}]{more2009}
More A.,  McKean J.~P.,  More S.,  Porcas R.~W.,  Koopmans L. V.~E.,    Garrett
  M.~A.,  2009, Monthly Notices of the Royal Astronomical Society, 394, 174

\bibitem[\protect\citeauthoryear{{Myers} \& {et al.}}{{Myers} \& {et
  al.}}{1995}]{myers1995}
{Myers} S.~T.,  {et al.} 1995, The Astrophysical Journal, 447, L5

\bibitem[\protect\citeauthoryear{{Nierenberg}, {Treu}, {Wright}, {Fassnacht} \&
  {Auger}}{{Nierenberg} et~al.}{2014}]{nierenberg2014}
{Nierenberg} A.~M.,  {Treu} T.,  {Wright} S.~A.,  {Fassnacht} C.~D.,    {Auger}
  M.~W.,  2014, ArXiv e-prints

\bibitem[\protect\citeauthoryear{Nishimichi, Shirata, Taruya, Yahata, Saito,
  Suto, Takahashi, Yoshida, Matsubara, Sugiyama, Kayo, Jing \&
  Yoshikawa}{Nishimichi et~al.}{2009}]{nishimichi2009}
Nishimichi T.,  Shirata A.,  Taruya A.,  Yahata K.,  Saito S.,  Suto Y.,
  Takahashi R.,  Yoshida N.,  Matsubara T.,  Sugiyama N.,  Kayo I.,  Jing
  Y.~P.,    Yoshikawa K.,  2009, Publications of the Astronomical Society of
  Japan, 61, 321

\bibitem[\protect\citeauthoryear{{Planck Collaboration}, {Ade}, {Aghanim},
  {Alves}, {Armitage-Caplan}, {Arnaud}, {Ashdown}, {Atrio-Barandela}, {Aumont},
  {Aussel} \& et al.}{{Planck Collaboration} et~al.}{2014}]{ade2014}
{Planck Collaboration} {Ade} P.~A.~R.,  {Aghanim} N.,  {Alves} M.~I.~R.,
  {Armitage-Caplan} C.,  {Arnaud} M.,  {Ashdown} M.,  {Atrio-Barandela} F.,
  {Aumont} J.,  {Aussel} H.,    et al. 2014, Astronomy and astrophysics, 571,
  A1

\bibitem[\protect\citeauthoryear{Ros, Guirado, Marcaide, Perez-Torres, Falco,
  Munoz, Alberdi \& Lara}{Ros et~al.}{2000}]{Ros2000}
Ros E.,  Guirado J.~C.,  Marcaide J.~M.,  Perez-Torres M.~A.,  Falco E.~E.,
  Munoz J.~A.,  Alberdi A.,    Lara L.,  2000, Astronomy and Astrophysics, 362,
  845

\bibitem[\protect\citeauthoryear{Schechter \& Moore}{Schechter \&
  Moore}{1993}]{schechter1993}
Schechter P.~L.,  Moore C.~B.,  1993, Astronomical Journal, 105, 1

\bibitem[\protect\citeauthoryear{{Schneider}, {Smith}, {Macci{\`o}} \&
  {Moore}}{{Schneider} et~al.}{2012}]{schneider2012}
{Schneider} A.,  {Smith} R.~E.,  {Macci{\`o}} A.~V.,    {Moore} B.,  2012,
  Monthly Notices of Royal Astronomical Society, 424, 684

\bibitem[\protect\citeauthoryear{{Schneider} \& {Sluse}}{{Schneider} \&
  {Sluse}}{2013}]{schneider2013}
{Schneider} P.,  {Sluse} D.,  2013, Astronomy and astrophysics, 559, A37

\bibitem[\protect\citeauthoryear{{Schultz}, {O{\~n}orbe}, {Abazajian} \&
  {Bullock}}{{Schultz} et~al.}{2014}]{schultz2014}
{Schultz} C.,  {O{\~n}orbe} J.,  {Abazajian} K.~N.,    {Bullock} J.~S.,  2014,
  Monthly Notices of Royal Astronomical Society, 442, 1597

\bibitem[\protect\citeauthoryear{{Seljak}, {Makarov}, {McDonald} \&
  {Trac}}{{Seljak} et~al.}{2006}]{seljak2006}
{Seljak} U.,  {Makarov} A.,  {McDonald} P.,    {Trac} H.,  2006, Physical
  Review Letters, 97, 191303

\bibitem[\protect\citeauthoryear{{Sluse}, {Chantry}, {Magain}, {Courbin} \&
  {Meylan}}{{Sluse} et~al.}{2012}]{sluse2012}
{Sluse} D.,  {Chantry} V.,  {Magain} P.,  {Courbin} F.,    {Meylan} G.,  2012,
  Astronomy and Astrophysics, 538, A99

\bibitem[\protect\citeauthoryear{{Smith} \& {Markovic}}{{Smith} \&
  {Markovic}}{2011}]{sm2011}
{Smith} R.~E.,  {Markovic} K.,  2011, Physical Review D, 84, 063507

\bibitem[\protect\citeauthoryear{Smith, Peacock, Jenkins, White, Frenk, Pearce,
  Thomas, Efstathiou \& Couchman}{Smith et~al.}{2003}]{smith2003}
Smith R.~E.,  Peacock J.~A.,  Jenkins A.,  White S. D.~M.,  Frenk C.~S.,
  Pearce F.~R.,  Thomas P.~A.,  Efstathiou G.,    Couchman H. M.~P.,  2003,
  Monthly Notices of the Royal Astronomical Society, 341, 1311

\bibitem[\protect\citeauthoryear{Springel}{Springel}{2005}]{springel2005}
Springel V.,  2005, Monthly Notices of the Royal Astronomical Society, 364,
  1105

\bibitem[\protect\citeauthoryear{Springel, Yoshida \& White}{Springel
  et~al.}{2001}]{springel2001}
Springel V.,  Yoshida N.,    White S. D.~M.,  2001, New Astronomy, 6, 79

\bibitem[\protect\citeauthoryear{Sugai, Kawai, Shimono, Hattori, Kosugi,
  Kashikawa, Inoue \& Chiba}{Sugai et~al.}{2007}]{sugai2007}
Sugai H.,  Kawai A.,  Shimono A.,  Hattori T.,  Kosugi G.,  Kashikawa N.,
  Inoue K.~T.,    Chiba M.,  2007, Astrophysical Journal, 660, 1016

\bibitem[\protect\citeauthoryear{{Suyu}, {Hensel}, {McKean}, {Fassnacht},
  {Treu}, {Halkola}, {Norbury}, {Jackson}, {Schneider}, {Thompson}, {Auger},
  {Koopmans} \& {Matthews}}{{Suyu} et~al.}{2012}]{suyu2012}
{Suyu} S.~H.,  {Hensel} S.~W.,  {McKean} J.~P.,  {Fassnacht} C.~D.,  {Treu} T.,
   {Halkola} A.,  {Norbury} M.,  {Jackson} N.,  {Schneider} P.,  {Thompson} D.,
   {Auger} M.~W.,  {Koopmans} L.~V.~E.,    {Matthews} K.,  2012, The
  Astrophysical Journal, 750, 10

\bibitem[\protect\citeauthoryear{{Takahashi} \& {Inoue}}{{Takahashi} \&
  {Inoue}}{2014}]{takahashi-inoue2014}
{Takahashi} R.,  {Inoue} K.~T.,  2014, Monthly Notices of Royal Astronomical
  Society, 440, 870

\bibitem[\protect\citeauthoryear{Takahashi, Sato, Nishimichi, Taruya \&
  Oguri}{Takahashi et~al.}{2012}]{takahashi2012}
Takahashi R.,  Sato M.,  Nishimichi T.,  Taruya A.,    Oguri M.,  2012,
  Astrophysical Journal, 761, 152

\bibitem[\protect\citeauthoryear{Tonry}{Tonry}{1998}]{tonry1998}
Tonry J.~L.,  1998, Astronomical Journal, 115, 1

\bibitem[\protect\citeauthoryear{Tonry \& Kochanek}{Tonry \&
  Kochanek}{1999}]{tonry1999}
Tonry J.~L.,  Kochanek C.~S.,  1999, Astronomical Journal, 117, 2034

\bibitem[\protect\citeauthoryear{{Viel}, {Becker}, {Bolton} \&
  {Haehnelt}}{{Viel} et~al.}{2013}]{viel2013}
{Viel} M.,  {Becker} G.~D.,  {Bolton} J.~S.,    {Haehnelt} M.~G.,  2013,
  Physical Review D, 88, 043502

\bibitem[\protect\citeauthoryear{{Viel}, {Lesgourgues}, {Haehnelt}, {Matarrese}
  \& {Riotto}}{{Viel} et~al.}{2005}]{viel2005}
{Viel} M.,  {Lesgourgues} J.,  {Haehnelt} M.~G.,  {Matarrese} S.,    {Riotto}
  A.,  2005, Physical Review D, 71, 063534

\bibitem[\protect\citeauthoryear{{Viel}, {Markovic}, {Baldi} \&
  {Weller}}{{Viel} et~al.}{2012}]{viel2012}
{Viel} M.,  {Markovic} K.,  {Baldi} M.,    {Weller} J.,  2012, Monthly Notices
  of Royal Astronomical Society, 421, 50

\bibitem[\protect\citeauthoryear{{Wang} \& {White}}{{Wang} \&
  {White}}{2007}]{wang2007}
{Wang} J.,  {White} S.~D.~M.,  2007, Monthly Notices of Royal Astronomical
  Society, 380, 93

\bibitem[\protect\citeauthoryear{{White} \& {Croft}}{{White} \&
  {Croft}}{2000}]{wc2000}
{White} M.,  {Croft} R.~A.~C.,  2000, Astrophysical Journal, 539, 497

\bibitem[\protect\citeauthoryear{Xu, Mao, Wang, Springel, Gao, White, Frenk,
  Jenkins, Li \& Navarro}{Xu et~al.}{2009}]{xu2009}
Xu D.,  Mao S.,  Wang J.,  Springel V.,  Gao L.,  White S.,  Frenk C.,  Jenkins
  A.,  Li G.,    Navarro J.,  2009, Monthly Notices of the Royal Astronomical
  Society, 398, 1235

\bibitem[\protect\citeauthoryear{{Xu}, {Sluse}, {Gao}, {Wang}, {Frenk}, {Mao},
  {Schneider} \& {Springel}}{{Xu} et~al.}{2014}]{xu2014}
{Xu} D.,  {Sluse} D.,  {Gao} L.,  {Wang} J.,  {Frenk} C.,  {Mao} S.,
  {Schneider} P.,    {Springel} V.,  2014, ArXiv e-prints

\bibitem[\protect\citeauthoryear{Xu, Mao, Cooper, Gao, Frenk, Angulo \&
  Helly}{Xu et~al.}{2012}]{xu2012}
Xu D.~D.,  Mao S.,  Cooper A.~P.,  Gao L.,  Frenk C.~S.,  Angulo R.~E.,
  Helly J.,  2012, Monthly Notices of the Royal Astronomical Society, 421, 2553

\bibitem[\protect\citeauthoryear{Xu, Mao, Cooper, Wang, Gao, Frenk \&
  Springel}{Xu et~al.}{2010}]{xu2010}
Xu D.~D.,  Mao S.~D.,  Cooper A.~P.,  Wang J.,  Gao L.~A.,  Frenk C.~S.,
  Springel V.,  2010, Monthly Notices of the Royal Astronomical Society, 408,
  1721

\end{thebibliography}
%\bibliography{test}

\end{document}